\documentclass[%
aps,
pre,
 amsmath,amssymb,
 superscriptaddress,
preprint,%
longbibliography
]{revtex4-2}

\usepackage[top=1in, bottom=1.25in, width=180mm]{geometry}
\usepackage{graphicx}% Include figure files
\usepackage{dcolumn}% Align table columns on decimal point
\usepackage{bm}% bold math
\usepackage{ulem}
\usepackage[utf8]{inputenc}
\usepackage[T1]{fontenc}
\usepackage{mathptmx}
\usepackage{textcomp} % for \textsuperscript, \textmu and many more
\usepackage{nicefrac}
\usepackage{xcolor}
\usepackage{gensymb}
\usepackage{multibib}
\usepackage{amsmath}
\usepackage{upgreek}
\usepackage{hyperref}
\usepackage{ragged2e}
\usepackage{physics}

\usepackage[activate={true,nocompatibility},final,tracking=true,kerning=true,spacing=true,factor=1100,stretch=10,shrink=10]{microtype}
\SetProtrusion{encoding={*},family={*},series={*},size={6,7}}
{1={ ,750},2={ ,500},3={ ,500},4={ ,500},5={ ,500},
6={ ,500},7={ ,600},8={ ,500},9={ ,500},0={ ,500}}
\SetExtraKerning[unit=space]
{encoding={*}, family={*}, series={*}, size={footnotesize,small,normalsize}}
{\textendash={400,400}, % en-dash, add more space around it
"28={150,100}, % left bracket, add space from right
"29={100,150}, % right bracket, add space from left
\textquotedblleft={ ,150}, % left quotation mark, space from right
\textquotedblright={150, }} % right quotation mark, space from left
\SetTracking{encoding={*}, shape=sc}{10}
\usepackage{lineno}
\newcommand{\ve}[1]{\boldsymbol{#1}}

\DeclareMathAlphabet{\pazocal}{OMS}{zplm}{m}{n}
\newcommand{\Tb}{\pazocal{T}}

\newcommand{\Da}{\mathcal{D}}
\newcommand{\Ea}{\mathcal{E}}
\newcommand{\figlabel}{Fig.~}
\newcommand{\figlabelext}{Extended Data Fig.~}

\usepackage[font={small, stretch=1.2}, figurename={Fig.}, labelfont={bf}, margin=0pt, justification=raggedright,format=plain]{caption}
\DeclareCaptionJustification{raggedright}{\justifying}
\usepackage{cleveref}

\begin{document}

\title{The role of non-equilibrium populations in dark exciton formation}

\author{Paul Werner} %
\affiliation{I. Physikalisches Institut, Georg-August-Universit\"at G\"ottingen, Friedrich-Hund-Platz 1, 37077 G\"ottingen, Germany}

\author{Wiebke Bennecke} %
\affiliation{I. Physikalisches Institut, Georg-August-Universit\"at G\"ottingen, Friedrich-Hund-Platz 1, 37077 G\"ottingen, Germany}

\author{Jan Philipp Bange} %
\affiliation{I. Physikalisches Institut, Georg-August-Universit\"at G\"ottingen, Friedrich-Hund-Platz 1, 37077 G\"ottingen, Germany}

\author{Giuseppe Meneghini} 
\affiliation{Fachbereich Physik, Philipps-Universit{\"a}t, 35032 Marburg, Germany}

\author{David Schmitt} %
\affiliation{I. Physikalisches Institut, Georg-August-Universit\"at G\"ottingen, Friedrich-Hund-Platz 1, 37077 G\"ottingen, Germany}

\author{Marco Merboldt} %
\affiliation{I. Physikalisches Institut, Georg-August-Universit\"at G\"ottingen, Friedrich-Hund-Platz 1, 37077 G\"ottingen, Germany}

\author{Anna M. Seiler} %
\affiliation{I. Physikalisches Institut, Georg-August-Universit\"at G\"ottingen, Friedrich-Hund-Platz 1, 37077 G\"ottingen, Germany}

\author{AbdulAziz AlMutairi} 
\affiliation{Department of Engineering, University of Cambridge, Cambridge CB3 0FA, U.K.}
\affiliation{King Fahd University of Petroleum \& Minerals (KFUPM), Dhahran, Saudi Arabia}

\author{Kenji Watanabe} %
\affiliation{Research Center for Electronic and Optical Materials, National Institute for Materials Science, 1-1 Namiki, Tsukuba 305-0044, Japan}

\author{Takashi Taniguchi} %
\affiliation{Research Center for Materials Nanoarchitectonics, National Institute for Materials Science,  1-1 Namiki, Tsukuba 305-0044, Japan}

\author{G.~S.~Matthijs~Jansen}
\affiliation{I. Physikalisches Institut, Georg-August-Universit\"at G\"ottingen, Friedrich-Hund-Platz 1, 37077 G\"ottingen, Germany}

\author{Junde Liu}
\affiliation{I. Physikalisches Institut, Georg-August-Universit\"at G\"ottingen, Friedrich-Hund-Platz 1, 37077 G\"ottingen, Germany}

\author{Daniel Steil} %
\affiliation{I. Physikalisches Institut, Georg-August-Universit\"at G\"ottingen, Friedrich-Hund-Platz 1, 37077 G\"ottingen, Germany}

\author{Stephan Hofmann} 
\affiliation{Department of Engineering, University of Cambridge, Cambridge CB3 0FA, U.K.}

\author{R. Thomas Weitz} %
\affiliation{I. Physikalisches Institut, Georg-August-Universit\"at G\"ottingen, Friedrich-Hund-Platz 1, 37077 G\"ottingen, Germany}
\affiliation{International Center for Advanced Studies of Energy Conversion (ICASEC), University of Göttingen, Göttingen, Germany}

\author{Ermin Malic} 
\affiliation{Fachbereich Physik, Philipps-Universit{\"a}t, 35032 Marburg, Germany}
\affiliation{Department of Physics, Chalmers University of Technology, Gothenburg, Sweden}

\author{Stefan Mathias} \email{smathias@uni-goettingen.de}%
\affiliation{I. Physikalisches Institut, Georg-August-Universit\"at G\"ottingen, Friedrich-Hund-Platz 1, 37077 G\"ottingen, Germany}
\affiliation{International Center for Advanced Studies of Energy Conversion (ICASEC), University of Göttingen, Göttingen, Germany}

\author{Marcel Reutzel} \email{marcel.reutzel@uni-marburg.de}%
\affiliation{I. Physikalisches Institut, Georg-August-Universit\"at G\"ottingen, Friedrich-Hund-Platz 1, 37077 G\"ottingen, Germany}
\affiliation{Fachbereich Physik, Philipps-Universit{\"a}t, 35032 Marburg, Germany}

\begin{abstract}

In two-dimensional transition metal dichalcogenide structures, the optical excitation of a bright exciton may be followed by the formation of a plethora of lower energy dark states. In these formation and relaxation processes between different exciton species, non-equilibrium exciton and phonon populations play a dominant role, but remain so far largely unexplored as most states are inaccessible by regular spectroscopies. Here, on the example of homobilayer 2H-MoS$_2$, we realize direct access to the full exciton relaxation cascade from experiment and theory. By measuring the energy- and in-plane momentum-resolved photoemission spectral function, we reveal a distinct fingerprint for dark excitons in a non-equilibrium excitonic occupation distribution. In excellent agreement with microscopic many-particle calculations, we quantify the timescales for the formation of a non-equilibrium dark excitonic occupation and its subsequent thermalization to 85~fs and 150~fs, respectively. Our results provide a previously inaccessible view of the complete exciton relaxation cascade, which is of paramount importance for the future characterization of non-equilibrium excitonic phases and the efficient design of optoelectronic devices based on two-dimensional materials.
\end{abstract}
\maketitle
\newpage

\begin{figure}[bt]
    \centering
    \includegraphics[width=\linewidth]{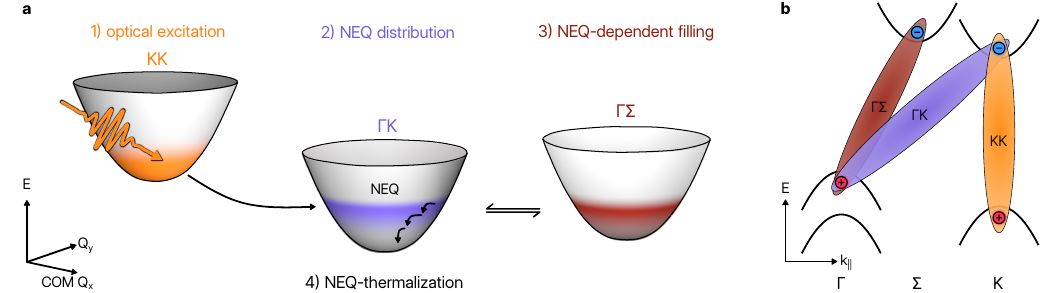}
    \caption{
	\textbf{Formation and thermalization of non-equilibrium excitonic occupations.}
    (a) The energy- and $\ve{Q}$-momentum-dispersive low energy landscape of excitons in homobilayer 2H-MoS$_2$ is composed of the optically bright KK exciton and two momentum-indirect $\Gamma$K and $\Gamma\Sigma$ excitons. Subsequent to the (1) optical excitation of KK excitons (orange laser pulse), intervalley hole-transfer occurs via exciton-phonon scattering and leads to the formation of a (2) NEQ $\Gamma$K exciton population (violett region). While this population is not yet thermalized, additional exciton-phonon scattering processes can (3) transfer population to the $\Gamma\Sigma$ dark exciton state. Because the NEQ exciton population in $\Gamma$K thermalizes, the $\Gamma$K-$\Gamma\Sigma$ scattering channel becomes gradually less efficient.
    (b) Corresponding single-particle energy landscape: The KK, $\Gamma$K, and $\Gamma\Sigma$ excitons are composed of single-particle Bloch states originating from the K-valley valence and conduction bands (orange), the $\Gamma$-valley valence band and the K-valley conduction band (violet), and the $\Gamma$-valley valence band and the $\Sigma$-valley conduction band (brown), respectively.
    } 
    \label{fig:fig1}
\end{figure}

Semiconducting transition metal dichalcogenides have become a promising platform for the next-generation optoelectronic devices~\cite{Liang20advmat, Mueller2018npj}. This is because the energy landscape of optical excitations is built up by Coulomb-bound electron-hole-pairs, so-called excitons~\cite{Wang18rmp,Perea22apl}, whose energy E$_{\rm exc}$, kinetic center-of-mass momentum (COM) $\ve{Q}$, and formation and thermalization dynamics can be controlled via multiple tuning knobs such as interlayer hybridization~\cite{Ruiz19prb,Wang17prb}, charge-carrier density~\cite{Chernikov15prl}, dielectric environment~\cite{Raja17natcomm}, moiré potential~\cite{Alexeev19nat,Tran19nat,Seyler19nat}, or the proximity of higher order correlated phases~\cite{Xu20nat}. With regard to the femto- to picosecond exciton formation and thermalization dynamics~\cite{Jin18natnano},
\figlabel\ref{fig:fig1}a illustrates how quasiparticle scattering processes can lead to the build-up of strong non-equillibrium exciton populations: After optical excitation of bright exciton states within the light-cone with $\ve{Q}\approx0$, subsequent exciton-phonon, exciton-exciton, and exchange-driven scattering events can lead to a redistribution of the energy- and $\ve{Q}$-momentum-dependent E$^i_{\rm exc}(\ve{Q})$ exciton population within the same exciton parabola ($i$=KK), and towards energetically favorable dark exciton states that can be of momentum-indirect or spin-forbidden nature ($i$=$\Gamma$K, $\Gamma\Sigma$)~\cite{Zhang15prl,Raja18nanolett,Jiang21sciadv,Selig182Dmat,Merkl19natmat,Madeo20sci,Wallauer21nanolett,Bange232DMaterials}. In all of these cases, the exciton and phonon populations are initially formed in a non-equilibrium (NEQ) condition ($\ve{Q}\neq0$)~\cite{Caruso21jpcl,Ovesen19comphys}, and additional energy dissipation processes are necessary for the exciton population to thermalize towards $\ve{Q}\approx0$. Therefore, to understand and tailor the excitonic formation and thermalization processes, an understanding of NEQ exciton dynamics is necessary.

While microscopic model calculations naturally capture the full E$^i_{\rm exc}(\ve{Q})$-resolved exciton dynamics~\cite{Selig182Dmat,Meneghini23acspho}, experimental access is challenging. This is because ultrafast all-optical spectroscopies are typically only sensitive to transitions within the light cone, and are fairly limited with regard to $\ve{Q}$-momentum-resolved experiments~\cite{Jin18natnano}. Still, selected ultrafast all-optical spectroscopies~\cite{Policht23natcom,Rosati20acsphotonics,Poellmann15natmat,Merkl19natmat} could differentiate the respective timescales of formation and thermalization, for example, by evaluating phonon-assisted secondary~\cite{Rosati20acsphotonics} or internal \textit{1s-2p}~\cite{Poellmann15natmat,Merkl19natmat} transitions. Complementary, the experimental capabilities of time- and angle-resolved photoemission spectroscopy (trARPES) might be ideally suited to resolve the E$^i_{\rm exc}(\ve{Q})$-resolved exciton dynamics~\cite{Reutzel24AdvPhysX}. Indeed, recent trARPES experiments enabled the direct experimental access to dark excitons whose wavefunction is composed of single-particle valence and conduction band states with different valley indices~\cite{Madeo20sci,Wallauer21nanolett,Dong20naturalsciences,Man21sciadv,Schmitt22nat,Karni22nat,kunin23prl,Bange232DMaterials,Bange24SciAdv,Schmitt25natpho,Bennecke24arxiv}. However, to probe the energy- and $\ve{Q}$-momentum-resolved dynamics, the scenario is more challenging, because the photoemission experiment is not sensitive to the exciton COM $\ve{Q}$, but to the in-plane momentum $\ve{k_{\rm ||}}$ of the photoemitted single-particle electrons that result from the break-up of excitons in the photoemission process. Hence, even though there are theoretical predictions on how $\ve{Q}$-momentum-dependent exciton populations would appear in the energy- and $\ve{k_{\rm ||}}$-momentum-resolved photoemission experiment~\cite{Perfetto16prb, Steinhoff2017natcom, Rustagi18prb, Christiansen19prb, Meneghini23acspho}, it has not yet been shown that trARPES can address the E$^i_{\rm exc}(\ve{Q})$-dependent dynamics of bright and dark excitons, and thus the full exciton relaxation cascade.

In this article, we demonstrate the contribution of NEQ excitonic occupations to formation and thermalization processes from experiment and theory. As a model system, we choose homobilayer 2H-MoS$_2$ (BL 2H-MoS$_2$), which exhibits a dark exciton landscape that enforces the contribution of non-equilibrium exciton distributions following an optical excitation (\figlabel\ref{fig:fig1}a,b)~\cite{Mak10prl,Scheuschner14prb,Meneghini23acspho}, and is thus ideally suited to create NEQ exciton populations. We follow the ultrafast scattering processes in experiment using time-resolved photoemission momentum microscopy~\cite{Keunecke20timeresolved,medjanik_direct_2017}, and in this way are able to get a clear signature of NEQ exciton occupations. Our data is supported by fully microscopic many-particle calculations based on density matrix formalism~\cite{Meneghini22naturalsciences,Schmitt22nat}, and we find that the NEQ distribution in homobilayer MoS$_2$ is formed with a rise time of $\approx 85$~fs and subsequently thermalizes with a time constant of $\approx 150$~fs. Moreover, we are able to illustrate that the energy- and $\ve{k_{\rm ||}}$-momentum-resolved photoemission spectral function indeed contains signatures of E$^i_{\rm exc}(\ve{Q})$-resolved exciton occupations in NEQ and in thermal equilibrium.
Therewith, our work opens the door to study NEQ exciton occupations and their role in the formation and thermalization dynamics in two-dimensional semiconductors. More broadly, this identification of specific NEQ  signatures in the photoemission spectral function can facilitate the future study of NEQ excitonic phases in 2D quantum materials and support the development of van-der-Waals semiconductor devices.

\section{The exciton energy landscape of homobilayer 2H-MoS$_2$}
Figrues~\ref{fig:fig1}a and \ref{fig:fig1}b illustrate the schematic low energy landscape of excitons in homobilayer 2H-MoS$_2$ in the exciton and the single-particle picture, respectively (ab initio calculated energy landscape in \Cref{tab:energylandscape}). The exciton energy landscape is composed of the optically bright A1s exciton and two additional momentum-indirect dark excitons in energetic proximity to each other. We label the A1s exciton as 'KK', because the exciton's hole- and electron-components reside in the K-valley valence and conduction bands, respectively (\figlabel\ref{fig:fig1}b). In the same notation, the two energetically relevant momentum-indirect dark excitons are labelled as $\Gamma$K and $\Gamma\Sigma$ excitons. From the schematic exciton energy landscape, it can be anticipated that the population of the optically excited KK excitons decays via intervalley hole-transfer (violet arrow) in an exciton-phonon scattering process to the energetically more favorable $\Gamma$K exciton state (\figlabel\ref{fig:fig1}a)~\cite{Raja18nanolett}. Here, it is important to emphasize that the energetic separation between the exciton band minima of the bright KK and the dark $\Gamma$K excitons in bilayer MoS$_2$ is $\approx400$~meV, and therefore much larger than typical phonon energies of $\approx30$~meV~\cite{Jin14prb}. Hence, exciton population transfer via exciton-phonon scattering from KK to $\Gamma$K must lead to a NEQ exciton distribution in the $\Gamma$K state, i.e. creation of NEQ dark excitons with considerable excess energy, as sketched in \figlabel\ref{fig:fig1}a by the violet region. It is this NEQ distribution and its subsequent thermalization towards $\ve{Q}\approx0$ that we aim to probe with time-resolved momentum microscopy.

Importantly, we can use the energetic proximity of the $\Gamma\Sigma$ exciton state to the $\Gamma$K exciton as a direct sensor for the NEQ exciton occupation. As long as a sufficiently non-thermal NEQ exciton distribution with $\ve{Q}\neq0$ exists in $\Gamma$K, exciton population can be transferred back and forth between the $\Gamma$K and $\Gamma\Sigma$ exciton states in phonon-mediated scattering events (double-sided arrows in \figlabel\ref{fig:fig1}a). As the NEQ distribution in $\Gamma$K thermalizes towards lower energies and $\ve{Q}\approx0$, this scattering channel is gradually suppressed. Therefore, when measuring the population dynamics in the $\Gamma$K and $\Gamma\Sigma$ state, we expect a distinct signature of the NEQ-dependent population dynamics.

\begin{figure}[bt!]
    \centering
    \includegraphics[width=\textwidth]{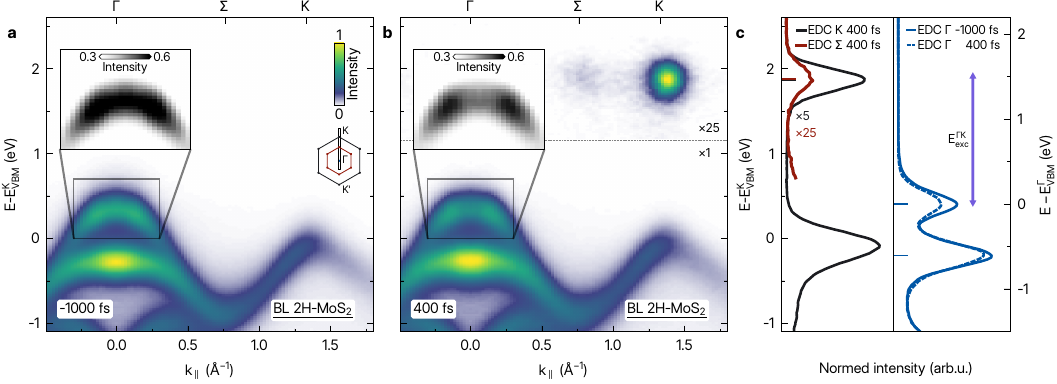}
    \caption{
	\textbf{Energy- and $\ve{k_{\rm ||}}$-momentum-resolved photoemission spectra of homobilayer 2H-MoS$_2$.}
    (a,b) Energy- and $\ve{k_{\rm ||}}$-momentum-resolved photoemission spectra collected along the $\Gamma$-$\Sigma$-K direction at delays of $\Delta t=-1000$~fs (a) and $\Delta t =400$~fs (b). In addition to the valence band structure at energies E-E$^{\rm K}_{\rm VBM}$<0.5~eV [(a) and (b)], photoemission spectral weight originating from the break-up of excitons is observed at the K and the $\Sigma$-valley for E-E$^{\rm K}_{\rm VBM}$>1~eV after optical excitation [(b) at 400 fs]. Additionally, at 400 fs, a distinct suppression of intensity at the $\Gamma$-valley valence band maximum indicates the presence of excitonic holes in 2H-MoS$_2$ [compare insets in (a) and (b)].
    (c) Energy-distribution-curves integrated in a 0.09$\times$0.09~\AA$^{-2}$ wide momentum-region around the K-valley (black), the $\Sigma$-valley (brown), and the $\Gamma$-valley (blue). Fitted peak positions are marked by colored horizontal lines.
    }
    \label{fig:fig2}
\end{figure}

We start our analysis with the identification of the different bright and dark exciton signatures in photoelectron spectra. To do so, we use our Göttingen time-, energy-, and momentum-resolved photoemission experiment that is based on the combination of a 500~kHz high-harmonic generation light source (26.5~eV, 20~fs, $p$-polarized) and a time-of-flight momentum microscope (details in methods)~\cite{medjanik_direct_2017,Keunecke20timeresolved}. Figure~\ref{fig:fig2}a shows an energy- and $\ve{k_{\rm ||}}$-momentum-resolved photoemission spectrum taken along the $\Gamma$-$\Sigma$-K direction at a pump-probe delay of $\Delta t=-1000$~fs (i.e., before the optical excitation). In agreement with earlier static ARPES experiments on homobilayer 2H-MoS$_2$~\cite{Wencan13prl}, we find the highest occupied valence band at the $\Gamma$-valley. The K-valley valence band maximum (VBM) is found at a lower energy ($E-E^{\Gamma}_{\rm VBM}=-0.38 \pm 0.05$~eV), clearly validating the indirect band gap of homobilayer 2H-MoS$_2$~\cite{Mak10prl,Wencan13prl}. 

Next, we focus on the energy- and $\ve{k_{\rm ||}}$-momentum-resolved photoemission spectrum collected at a pump-probe delay of $\Delta t = 400$~fs (\figlabel\ref{fig:fig2}b, 1.9~eV, 40~fs, $s$-polarized, 3~$\upmu$J/cm$^{-2}$ absorbed fluence, 9$\times10^{12}$~cm$^{-2}$ exciton density, see methods). At this pump-probe time delay, we find additional photoemission spectral weight at (i) the K-valley (E-E$^{\rm K}_{\rm VBM}=1.90\pm 0.05$~eV), (ii) the $\Sigma$-valley (E-E$^{\rm K}_{\rm VBM,K}=1.92\pm 0.05$~eV), and, furthermore, (iii) a distinct suppression of photoemission spectral weight at the $\Gamma$-valley VBM (insets in \figlabel\ref{fig:fig2}a,b). We can attribute these photoemission signatures to specific exciton species in homobilayer 2H-MoS$_2$ as follows: The K and $\Sigma$-valley photoemission spectral weight results from the photoemitted electrons that did belong to excitons whose electron-component resided in the respective K and $\Sigma$-valleys~\cite{Madeo20sci,Wallauer21nanolett,Dong20naturalsciences,Schmitt22nat,Reutzel24AdvPhysX}. Hence, the K-valley spectral weight is indicative for the presence of KK and/or $\Gamma$K excitons (cf. \figlabel\ref{fig:fig1}b). Similarly, the $\Sigma$-valley spectral weight results from the break-up of $\Gamma\Sigma$ excitons
. Intriguingly, the $\Gamma$K and $\Gamma\Sigma$ exciton's hole contributions can also directly be verified by the observation of missing valence band spectral weight at the VBM of the $\Gamma$-valley (insets in \figlabel\ref{fig:fig2}a,b, compare with Ref.~\cite{Karni22nat}). Finally, we evaluate the energies E$_{\rm exc}^i$ of all contributing bright and dark excitons (i~=~KK, $\Gamma$K, $\Gamma\Sigma$, \figlabel\ref{fig:fig2}c). As detailed in the methods, we do this by considering the conservation of energy and momentum as the two-particle exciton is broken during the photoemission process~\cite{Reutzel24AdvPhysX,Bange24SciAdv,Bennecke24natcom}.
The experimentally quantified energy landscape of all contributing excitons is summarized in \Cref{tab:energylandscape}, and, importantly, shows an excellent agreement with the ab initio calculated exciton energies and earlier photoluminescence experiments~\cite{Mak10prl,Scheuschner14prb}.

\begin{table}[!h]
    \centering
    \caption{Low energy landscape of excitons as extracted in with trARPES, theory, and photoluminescence (PL)~\cite{Mak10prl,Scheuschner14prb}.}
    \label{tab:energylandscape}
    \begin{tabular}{c||c|c|c}
        \hline
         Exciton & trARPES: $E_{\rm exc}^{i}$ (eV)   & Theory: $E_{\rm exc}^{i}$ (eV) & PL (eV)~\cite{Mak10prl,Scheuschner14prb}\\\hline\hline
         KK             & 1.90$\pm$0.05   & 1.89 & 1.88 \\
         K$\Sigma$      & --              & 1.98 & -\\
         $\Gamma$K      & 1.52$\pm$0.05   & 1.50 & 1.5\\
         $\Gamma\Sigma$ & 1.54$\pm$0.05   & 1.55 & - \\\hline
    \end{tabular}
\end{table}

\section{Signatures of NEQ exciton occupations in the ultrafast dynamics}

\begin{figure}[bt]
    \includegraphics[width=\textwidth]{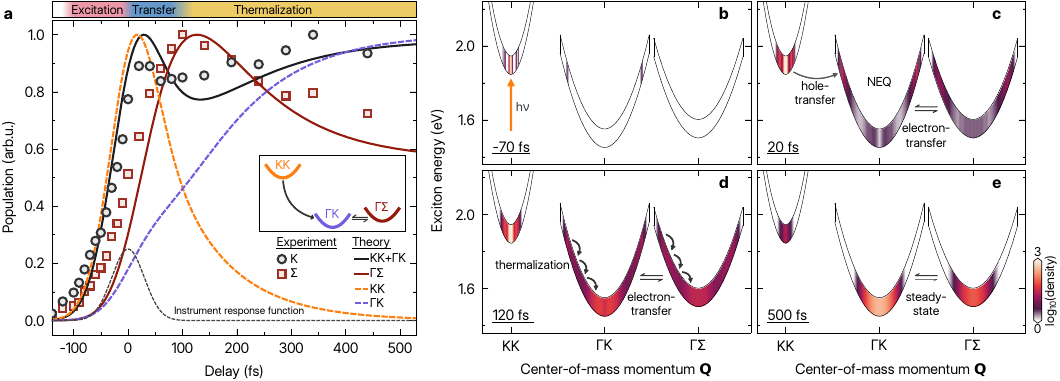}
    \caption{\textbf{Exciton formation and thermalization dynamics in homobilayer 2H-MoS$_2$.}
    (a) Energy- and $\ve{k_{\rm ||}}$-momentum-integrated excitonic photoemission spectral weight extracted at the K-valley (black circles) and the $\Sigma$-valley (red squares) as a function of pump-probe delay. At $\approx100$~fs, where the $\Sigma$-valley spectral weight is maximum, the K valley signal shows a peculiar minimum followed by a subsequent second rise of intensity. The experimentally measured dynamics are well described by energy- and $\ve{Q}$-momentum-integrated microscopic model calculations that capture the NEQ population dynamics of KK (orange), $\Gamma$K (violet), and $\Gamma\Sigma$ (brown) excitons, and where the K-valley ($\Sigma$-valley) spectral weight corresponds to the sum of the KK and $\Gamma$K populations (the $\Gamma\Sigma$ population). The exciton dynamics can be grouped into the delay-regimes of optical excitation (-100~fs<$\Delta t$<0~fs, red), intervalley charge-transfer and dark exciton formation (0~fs<$\Delta t$<100~fs, blue), and the thermalization of the exciton populations towards a $\Gamma$K$\rightleftharpoons\Gamma\Sigma$ steady state  ($\Delta t$>100~fs, yellow). The inset illustrates the exciton cascade KK$\rightarrow\Gamma$K$\rightleftharpoons\Gamma\Sigma$.
    (b-e) Delay-dependent snapshots of the calculated energy- and $\ve{Q}$-momentum-resolved exciton population dynamics for KK, $\Gamma$K, and $\Gamma\Sigma$ excitons subsequent to the resonant excitation of KK excitons. Exciton-phonon scattering leads to the formation of a NEQ distribution of KK, $\Gamma$K, and $\Gamma\Sigma$ excitons (b,c). Only subsequent intra- and intervalley scattering processes lead to the thermalization of the exciton population distribution and the establishment of a $\Gamma$K$\rightleftharpoons\Gamma\Sigma$ steady state (d,e).
    }
    \label{fig:fig3}
\end{figure}

After the identification of the different bright and dark excitons in the photoelectron spectra, we now consider the pump-probe delay-resolved experiment to measure NEQ-dependent population transfer between the excitonic states. Specifically, since the photoemission signatures of the KK and $\Gamma$K exictons coincide, we expect unusual dynamics for photoelectrons detected at the K-point, as two different exciton species contribute to the photoemission intensity. Figure~\ref{fig:fig3}a shows the pump-probe delay-dependent total photoelectron intensity at the K-valley (black circles) and the $\Sigma$-valley (red squares), which corresponds to the sum of the population dynamics of KK and $\Gamma$K excitons (signal at K) and $\Gamma\Sigma$ excitons (signal at $\Sigma$). As can directly be seen from the data, the signal at K (black circles) shows a pronounced dip around 100~fs and a delayed rise thereafter, which would not be expected for a simple cascaded population transfer. Indeed, this dip can only be understood if a NEQ population transfer from $\Gamma$K to $\Gamma\Sigma$ excitons is considered: while the population at $\Gamma$K is in a strong NEQ, excitons that transfer from $\Gamma$K to $\Gamma\Sigma$ do not contribute to spectral weight at the K-valley.  
Simultaneously, photoemission spectral weight at the $\Sigma$ valley increases as $\Gamma\Sigma$ exciton occupation is formed. At later times, the drop of $\Sigma$ valley spectral weight and the second rise at K is then caused by the back-transfer from $\Gamma\Sigma$ to $\Gamma$K due to thermalization of the NEQ distribution.

In order to support this experimental proposition more quantitatively, we simulate the delay-, energy-, and $\ve{Q}$-momentum-dependent population dynamics of KK, $\Gamma$K, and $\Gamma\Sigma$ excitons on a microscopic footing with the exciton density matrix formalism considering exciton-light and exciton-phonon interactions
~\cite{Meneghini22naturalsciences,Schmitt22nat} (see methods). The calculated energy- and $\ve{Q}$-momentum-integrated population dynamics of all contributing excitons is overlayed as lines on the experimental data in \figlabel\ref{fig:fig3}a (normalized to match experiment, non-normalized theory curves in \figlabelext\ref{fig:theory_time_traces_no_norm}). Importantly, we find that the dip in the experimental K-valley dynamics is also present in our theoretical analysis, if we consider the joint occupation dynamics of both excitons that are probed at the K valley (i.e., KK and $\Gamma$K excitons, solid black line in \figlabel\ref{fig:fig3}a). Again, we emphasize that this distinct feature in the dynamics is caused by NEQ exciton population that is transferred from $\Gamma$K (blue line) to $\Gamma\Sigma$ (brown line). Hence, our direct comparison of momentum microscopy data and microscopic theory provides so far unmatched quantitative access to the full relaxation cascade of bright and dark excitons in two-dimensional semiconductors.

Next, for a more detailed understanding of the role of the NEQ dynamics and the different timescales of exciton formation and thermalization, \figlabel\ref{fig:fig3}b-e shows selected snapshots of the calculated energy- and $\ve{Q}$-momentum-resolved exciton populations. The first interesting observation is that subsequent to the optical excitation of bright KK excitons, intravalley exciton-phonon scattering leads to a broadening of the KK exciton population to $\ve{Q}$-momenta outside the light-cone (cf. -70~fs and 20~fs data in \figlabel\ref{fig:fig3}b,c). In addition, the emission of K-phonons and M-phonons ($\Sigma$-phonons) enables intervalley hole transfer processes~\cite{Raja18nanolett} that transfer KK exciton population to $\Gamma$K, and intervalley electron transfer processes that transfer $\Gamma$K exciton population to the $\Gamma\Sigma$ state, respectively. We note that the direct transfer from KK to $\Gamma\Sigma$ is orders of magnitude less efficient, because it involves simultaneous electron- and hole-transfer events in a two-phonon scattering process and this therefore neglected~\cite{Meneghini23acspho}. 

In \figlabel\ref{fig:fig3}c, we can see that not only the initial $\Gamma$K, but also the $\Gamma\Sigma$ exciton population is clearly in a NEQ condition. Figure~\ref{fig:fig3}d at 120 fs shows that subsequent intra- and intervalley exciton-phonon scattering events lead to the thermalization of the exciton populations and the formation of a $\Gamma$K$\rightleftharpoons\Gamma\Sigma$ steady state (\figlabel\ref{fig:fig3}d, 120~fs). As the pump-probe delay evolves, the major part of the exciton population becomes of $\Gamma$K nature (\figlabel\ref{fig:fig3}e at 500 fs), because the $\Gamma$K exciton state is energetically lower by 50~meV compared to the $\Gamma\Sigma$ exciton state (see \figlabelext\ref{fig:theory_time_traces_no_norm}).
In particular, we find that the dynamics can be grouped into the delay-regimes of optical excitation (-100~fs < $\Delta t$ < 0~fs, red $\Delta t$-regime in \figlabel\ref{fig:fig3}a), intervalley charge-transfer and dark exciton formation (0~fs < $\Delta t$ < 100~fs, blue), and, finally, the thermalization of the exciton populations ($\Delta t$ > 100~fs, yellow). In the following, we examine how far these more complex dynamics can be captured via distinct spectroscopic fingerprints in the momentum microscopy experiment.

\section{Signatures of NEQ exciton occupations in the photoemission spectral function}

To capture the NEQ dynamics and the distinctly different timescales for exciton formation and thermalization, it is necessary to differentiate $\ve{Q}\approx0$ and $\ve{Q}\neq0$ exciton momentum distributions in the energy- and $\ve{k_{\rm ||}}$-momentum-resolved photoemission experiment. It is by now established that excitons with $\ve{Q}\approx0$ lead to hole-like photoemission signatures mimicking the $\ve{k_{\rm ||}}$-momentum-dependent valence band dispersion~\cite{Man21sciadv,Dong20naturalsciences}. For thermal and NEQ $\ve{Q}\neq0$ distributions, however, so far no experimental signatures have been discussed in detail. In a theoretical analysis, Rustagi and Kemper~\cite{Rustagi18prb} predicted that NEQ distributions ($\ve{Q}\neq0$) should lead to distinct deviations from the hole-like dispersion, including a broadening of the photoemission spectral weight on the photoelectron energy- and $\ve{k_{\rm ||}}$-momentum-axis~\cite{kunin23prl}. 

\begin{figure}[bt]
    \centering
    \includegraphics[width=\textwidth]{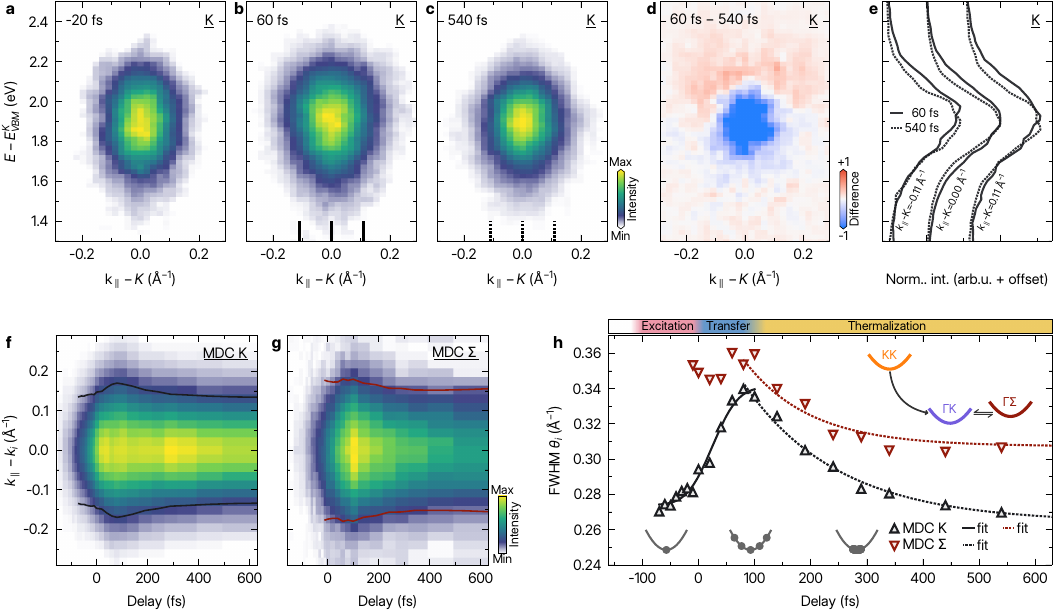}
    \caption{\textbf{Identification of spectroscopic fingerprints for NEQ and thermalized exciton populations in the energy- and $\ve{k_{\rm ||}}$-momentum-resolved photoemission spectral function.}
    (a,b,c) The shape of the photoemission snapshots taken at the K-valley show a distinct broadening on the energy- and $\ve{k_{\rm ||}}$-momentum-axis as the delay is varied from -20~fs, to 60~fs, and to 540~fs. 
    (d,e) In direct comparison of the 60~fs and 540~fs measurement, the energy- and $\ve{k_{\rm ||}}$-momentum-broadening becomes most evident in the difference map (d) and in momentum-filtered energy-distribution-curves (e).
    (f,g) Delay-dependent change of energy-integrated momentum-distribution-curves (MDCs) filtered at the K-valley (f) and the $\Sigma$-valley (g) (energy-integration-range: 2.4 to 1.5~eV). The solid lines indicate the FWHM $\theta_{\rm K}$ and $\theta_{\rm \Sigma}$ as extracted by fitting the MDCs with Gaussian distributions (see methods).
    (h) The peculiar increase and subsequent decrease of $\theta_{\rm K}$ (black triangles) and also the decrease of $\theta_{\rm \Sigma}$ (brown triangles) are indicative for the contribution of NEQ excitonic occupations to the photoemission spectral weight. The gray parabolas and dots in the bottom of the panel illustrate the contribution of $\ve{Q}\approx0$ excitons during optical excitation (-100~fs<$\Delta t$<0~fs, red delay-regime), the formation of $\ve{Q}\neq0$ excitonic occupations due to exciton-phonon scattering ((0~fs<$\Delta t$<100~fs, blue), and the subsequent thermalization towards $\ve{Q}\approx0$ ($\Delta t$>100~fs, yellow). 
    }
    \label{fig:fig4}
\end{figure}

Guided by the clear experimental and theoretical results of NEQ dynamics above, \figlabel\ref{fig:fig4}a-c illustrates E($\ve{k_{\rm ||}}$)-resolved photoemission snapshots at the K-valley taken at selected pump-probe delays. While changes in the broadening in energy and $k_{\rm ||}$-momentum are subtle in comparison to the experimental resolution and the overall width of the peak in energy and momentum, differences are clearly identifiable, in particular in the difference map shown in \figlabel\ref{fig:fig4}d and in the extracted momentum-filtered energy-distribution-curves in \figlabel\ref{fig:fig4}e. 
In the following, we show that this transient energy- and $\ve{k_{||}}$-momentum-broadening is a clear signature of NEQ exciton occupation dynamics.

To quantify the spectral changes more systematically, \figlabel\ref{fig:fig4}f,g show the delay-dependence of energy-integrated momentum-distribution-curves (MDCs) of the exciton photoemission signatures at the K (\figlabel\ref{fig:fig4}f) and the $\Sigma$ (\figlabel\ref{fig:fig4}g) valley. On-top of the color-coded photoemission yield, the black and brown lines indicate the full-width-at-half-maximum $\theta_{\rm i}$ (i~=~K,~$\Sigma$) obtained by Gaussian fits to the MDCs (see methods). Moreover, the fit results $\theta_{\rm i}$ are plotted as a function of pump-probe delay in \figlabel\ref{fig:fig4}h. We find distinct dynamics on the few-100~fs timescale: First, for the K-valley spectral weight, we find an increase of $\theta_{\rm K}$ that peaks at $\approx$100~fs and is followed by a subsequent decrease. Second, at the $\Sigma$-valley, we observe that $\theta_{\rm \Sigma}$ is nearly constant up to delays of $\approx$100 fs, before it decreases and saturates on a longer timescale. These observations can now directly be connected to the formation dynamics of dark NEQ excitonic occupations and their subsequent thermalization.

During temporal overlap of the pump- and probe laser-pulses, KK excitons are optically excited within the light-cone (red $\Delta t$-regime in \figlabel\ref{fig:fig4}h), so that the $\ve{Q}$-momentum-width of the excitation is initially small~\cite{Rustagi18prb}. This is also observed in experiment by a comparably small $\theta_{\rm K}$ value (\figlabel\ref{fig:fig4}h). 
For increasing pump-probe delay, the K valley MDC width $\theta_{\rm K}$ is expected to increase for two reasons: First, photoelectrons with slightly larger $\ve{k_{||}}$ are detected because of intravalley exciton-phonon scattering that induces a comparably small $\ve{Q}> 0$ KK distribution (\figlabel\ref{fig:fig3}b,c). Second, intervalley hole transfer leads to a strong $\ve{Q}> 0$ $\Gamma$K exciton population that, in consequence, must strongly contribute to the measured broadening at K (cf. \figlabel\ref{fig:fig3}c). Indeed, both the initial increase in $\theta_{\rm K}$ due to $\ve{Q}> 0$ KK exciton generation (red $\Delta t$-regime in \figlabel\ref{fig:fig4}h), and the additional stronger broadening due to the generation of the NEQ $\Gamma$K exciton population is visible in our data (blue $\Delta t$-regime). 

In addition to the NEQ exciton population dynamics in the KK and $\Gamma$K states, we can also analyze the formation and thermalization dynamics of the NEQ exciton population in the $\Gamma\Sigma$ state by carrying out the same analysis for the photoemission signal in the $\Sigma$-valley. 
Since exciton-phonon scattering also leads to a NEQ $\Gamma\Sigma$ exciton occupation, we expect to find a broad momentum distribution that gradually sharpens with time. Indeed, for delays $\Delta t$ as early as the signal-to-noise ratio allows to extract $\theta_{\rm \Sigma}$, we find that $\theta_{\rm \Sigma}$ is already at $\theta_{\rm \Sigma}(0~fs)\approx 0.35$~\AA$^{-1}$, while the signal in the K-valley is narrower with $\theta_{\rm K}(0~fs)\approx 0.29$~\AA$^{-1}$ due to the contribution of the narrower momentum distribution of KK-excitons in this valley. Hence, this observation confirms that 
the population for $\Gamma\Sigma$ excitons also starts in a strong NEQ distribution with $\ve{Q}> 0$, as predicted by theory. 

On the longer timescale, both NEQ exciton populations at $\Gamma$K and $\Gamma\Sigma$ are expected to thermalize via additional exciton-phonon scattering events that dissipate excess energy and lead towards a population with $\ve{Q}\approx0$. Indeed, \figlabel\ref{fig:fig4}h confirms that $\theta_{\rm K}$ and $\theta_{\rm \Sigma}$ decrease on timescales >100~fs (yellow $\Delta t$-regime), and we extract decay constants of $\tau = (170\pm15)$~fs and $\tau = (120\pm30)$~fs, respectively, by fitting single-exponential functions (dotted lines in \figlabel\ref{fig:fig4}h, \Cref{tab:time_scales}). Interestingly, the similar dynamics of the narrowing of the K and $\Sigma$-valley MDC width supports the theoretical prediction that not only intravalley exciton-phonon scattering thermalizes the NEQ distributions, but that phonon emission events mediate efficient intervalley electron-transfer between $\Gamma$K and $\Gamma\Sigma$ excitons. Hence, the excitonic occupation jointly thermalizes towards a $\Gamma$K$\rightleftharpoons\Gamma\Sigma$ steady state.

\section{Discussion and Outlook}

In summary, we have shown that the time-, energy-, and $\ve{k_{\rm ||}}$-momentum-resolved photoemission experiment not only provides access to bright (momentum-direct) and dark (momentum-indirect) excitons, but also to the energy- and $\ve{Q}$-momentum-resolved NEQ exciton population dynamics. Specifically, in homobilayer 2H-MoS$_2$, we have illustrated that we can directly access the formation dynamics of a NEQ exciton population of momentum-indirect $\Gamma$K and $\Gamma\Sigma$ excitons. We further showed how intra- and intervalley exciton-phonon scattering processes lead to the thermalization of the NEQ exciton population, where a steady state $\Gamma$K$\rightleftharpoons\Gamma\Sigma$ is established that subsequently decays on the picosecond timescale ($3.5\pm 0.3$~ps decay constant, \figlabelext\ref{fig:time_fits}).

More broadly, our work highlights that ultrafast momentum microscopy 
gives experimental access to the $\ve{Q}$-momentum-coordinate of optically bright and dark excitons. Therewith, it will be possible to study, e.g., the impact of other $\ve{Q}$-momentum-conserving scattering mechanisms such as Auger recombination~\cite{Poellmann15natmat} and exchange scattering~\cite{kunin23prl,Jiang21sciadv}. Moreover, we emphasize that such direct access to NEQ exciton dynamics is important for the study of emergent moiré-trapped exciton states~\cite{Alexeev19nat,Tran19nat,Seyler19nat,Karni22nat}, and for access to even more complex quasiparticles such as self-hybridized exciton-polaritons~\cite{Fei16prb} or even exciton-polariton condensates~\cite{Byrnes14natphys}.

\clearpage
\section{Methods}
\noindent\textbf{Sample Fabrication}

The sample was prepared by mechanically exfoliating a homobilayer of 2H-MoS$_2$ from a bulk crystal onto a PDMS substrate. The flake is then placed onto is placed on $\approx40$~nm thick hexagonal boron nitride (hBN) on a Niobium-doped Strontiumtitanite substrate (Nb-STO, 0.1~\% doping) via dry-transfer stamping \cite{seiler2024arxiv}.
A real space image of the sample acquired by photoemission electron microscopy is shown in \figlabelext\ref{fig:sample_img} with the position of the aperture used for the ARPES measurements marked as a black dashed line.
To prepare the sample for experiments, it is annealed at 670~K for 3~hours under ultrahigh vacuum conditions.
High sample quality is confirmed by the presence of clearly visible bands in the ARPES measurements with minimal background intensity.
\vspace{1cm}

\noindent\textbf{Femtosecond momentum microscopy}

A detailed description of the ultrafast momentum microscopy setup and its application to exfoliated van-der-Waals materials is provided in Refs.~\cite{Keunecke20timeresolved} and \cite{Schmitt22nat,Bange24SciAdv,Schmitt25natpho}, respectively. In short, the multi-dimensional photoemission data are acquired with a time-of-flight momentum microscope (ToF-MM, Surface Concept GmbH)~\cite{medjanik_direct_2017}, and the extreme-ultraviolet light pulses are generated with a table-top high harmonic generation beamline driven by a 300~W fiber laser (AFS Jena) operating at 500~kHz~\cite{Keunecke20timeresolved, Duvel22nanolett,Merboldt25NatPhys}. The pump photons are generated using an optical parametric amplifier (Light Conversion).

In all experiments reported in the manuscript, we measure energy- and in-plane momentum-resolved photoemission data cubes for selected pump-probe delays. The cross-correlation of the pump and probe laser pulses is quantified to 75$\pm$5~fs by evaluating the delay-dependence of the laser assisted photoelectric effect~\cite{miaja2006laser,Merboldt25NatPhys}. Before the data analysis, as detailed in Refs.~\cite{Schmitt22nat,Bange24SciAdv}, we correct the multi-dimensional data for space-charge and photovoltage effects~\cite{schonhense_multidimensional_2018,Roth24prl}. 
\vspace{1cm}

\noindent\textbf{Estimation of exciton densities}\\

To determine the density of optically excited excitons in the sample, the absorbed fluence has to be estimated. The spot size of the pump beam is determined in a real-space-resolved photoemission electron microscopy (PEEM) image to $215\times110$~µm$^2$. With the work function of the sample and the used photon energy (1.9~eV), photoemission occurs in non-linear order of three. Hence, the actual pump spot is larger by a factor of $\sqrt{3}$. With the incident pump power of 25~mW, we calculate the incident pump fluence 180~µJ/cm$^2$ (500~kHz repitition rate). Since the incident angle of the $s$-polarized pump photons in 22$^\circ$, following Fresnel equations, 26.3\% percent of the incident fluence is transmitted into the sample. A single monolayer MoS$_2$ absorbs 7.3\% of the incident light flux for 1.9~eV photon energy~\cite{Li14prb}. Assuming that both layers absorb an equal amount of photons and using the substrate-dependent correction factor from ref.~\cite{Li14prb} for the underlying hBN, the homobilayer absorbs approximately 6\% of the transmitted fluence.
This results in an exciton density of $9\cdot10^{12}$~cm$^{-2}$.
\vspace{1cm}

\noindent\textbf{Quantification of exciton energies $E_{\rm exc}^i$ with momentum microscopy}

Considering the conservation of energy and momentum during the break-up of excitons, momentum microscopy can facilitate the experimental characterization of the energy landscape of bright and dark excitons~\cite{Reutzel24AdvPhysX}. The energy $E_{\rm elec}$ of the measured single-particle photoelectron resulting from the break-up of excitons is given by $E_{\rm elec}=E_{\rm hole}+E^i_{\rm exc}+\hbar\omega$ ($\hbar\omega$: probe photon energy)~\cite{Weinelt04prl}. Here, $E_{\rm elec}$ is referenced to the energy of the hole, $E_{\rm hole}$, and therefore to the corresponding VBM in the specific valley, where the former exciton's hole-component resides after the breakup of the exciton by the photoemission process ($\hbar\omega$: probe photon energy). For example, for $\Gamma$K excitons, this implies that E$_{\rm exc}^{\rm \Gamma K}$ can be extracted by calculating the energy difference between the $\Gamma$-valley VBM and the K-valley exciton photoemission signal (\figlabel\ref{fig:fig2}c, violet double-headed arrow). The experimentally quantified energy landscape of all contributing excitons is summarized in \Cref{tab:energylandscape}, and, importantly, shows an excellent agreement with the ab initio calculated exciton energies and earlier photoluminescence experiments~\cite{Mak10prl,Scheuschner14prb}.
\vspace{1cm}

\noindent\textbf{Quantification of the momentum-width $\theta_i$ of the excitonic photoemission signal}

The energy- and center-of-mass momentum-resolved excitonic occupation is encoded in the in-plane momentum-resolved width of the photoemission intensity~\cite{Rustagi18prb}. We quantify this in-plane momentum-width $\theta_i$ for the K and $\Sigma$ valley excitonic photoemission spectral weight as follows: First, the ($k_x$,$k_y$)-momentum-resolved spectral weight is integrated in energy in a range from 1.5 to 2.4~eV for all six K or $\Sigma$ valleys. Second, the momentum maps are fitted with a 2D-Gaussian distribution with constant background. 
\vspace{1cm}

\noindent\textbf{Quantification of the timescales for exciton formation, thermalization, and decay}

\begin{table}[!pb]
\caption{Overview of the fit-results for the rise and decay of K and $\Sigma$ valley excited state spectral weight, as well as the rise and decay of the momentum-width $\theta_{\rm K}$ and $\theta_{\rm \Sigma}$.}
\label{tab:time_scales}
    \begin{tabular}{l|cc}
         & K valley intensity & $\Sigma$ valley intensity \\\hline\hline
       $\sigma_{\rm rise}$ (fs)       & 46$\pm$3      & 57$\pm$2     \\
       $\chi_{\rm rise}$ (fs)         & 24$\pm$2      & 65$\pm$2     \\\hline
       $\tau_{\rm decay}$ (fs)        & 3300$\pm$200  & -            \\
       $\tau_{\rm fast}$ (fs)         & -             & 290$\pm$50   \\
       $\tau_{\rm slow}$ (fs)         & -             & 3600$\pm$200 \\\hline
         & K valley FWHM & $\Sigma$ valley FWHM \\\hline\hline
       $\sigma_{\rm rise, \theta}$ (fs)& 36$\pm$7      & -            \\
       $\chi_{\rm rise, \theta}$ (fs)  & 25$\pm$5      & -            \\\hline
       $\tau_{\rm decay, \theta}$ (fs) & 170$\pm$13    & 120$\pm$30 
    \end{tabular}
\end{table}
Table~1 summarizes the timescales for exciton formation, thermalization, and decay, as directly quantified from experiment. Thereby, we evaluate not only the photoemission spectral weight at the K and the $\Sigma$ valley, but also the momentum-width $\theta_i$ ($i={\rm K,\Sigma}$) of the excitonic photoemission signatures. The respective fits are shown in \figlabelext\ref{fig:time_fits} and \figlabel\ref{fig:fig4}h.

The rise time of the K and $\Sigma$ valley spectral weight and also the rise time of the momentum-width $\theta_{\rm K}$ is fit with the error function
\begin{equation}
  f = \frac{A}{2}\cdot\left(1+\mathrm{erf}\left(\frac{x-\chi}{\sqrt{2}\sigma}\right)\right).
  \label{eq:rise_fit}
\end{equation}
Here, the fit parameters are the amplitude $A$, the position offset $\chi$, and the delay width $\sigma$. Note that the quantity referred to in the main text as rise time is the full-width-at-half-maximum of the of the error functions (i.e., $2.355\sigma$). The lifetime of the K and $\Sigma$ valley spectral weight and also the decreasing momentum-width $\theta_{\rm K}$ and $\theta_{\rm \Sigma}$ are fitted with single- or double-exponential functions of the form
\begin{equation}
  f(x) = A\cdot\exp\left(-\frac{x-\chi}{\tau}\right),
  \label{eq:exp_decay_single}
\end{equation}
or
\begin{equation}
  f(x) = A_1\cdot\exp\left(-\frac{x-\chi}{\tau_1}\right) + A_2\cdot\exp\left(-\frac{x-\chi}{\tau_2}\right).
  \label{eq:exp_decay_double}
\end{equation}
Here, the fit parameters of two different amplitudes $A_1$ and $A_2$, and timescales $\tau_1$, and $\tau_2$, and the delay offset $\chi$.
\vspace{1cm}

\noindent\textbf{Theoretical model}\\

We introduce the microscopic model adopted in the main text to track the energy-, center-of-mass-momentum- and time-resolved relaxation of excitons in the TMD homobilayer.
The Hamilton operator of the bilayer system, including the tunnelling term coupling the two layers, can be written as follows 
\begin{equation}\label{eq:4}
    H = H_0 + H_T = 
    \sum_{\mu,  \textbf{Q}}{ E^{\mu }_{\textbf{Q}} X^{\mu\dagger}_{\textbf{Q}} X^{\mu}_{\textbf{Q}} } + \sum_{\mu,\nu,\textbf{Q}}{ \Tb_{\mu \nu}  {X^{\mu\dagger }_{{\textbf{Q}}}X^{\nu}_{{\textbf{Q}} }}}
\end{equation}
with the superindex $\mu = (n^\mu, \zeta^\mu_e,\zeta^\mu_h,L^\mu_e,L^\mu_h)$ describing the exciton degrees of freedom, where $n$ is associated to the series of Rydberg-like states determining the relative electron-hole motion.
Moreover,  $\zeta_{e/h}$ and $L_{e/h}$ denote the electron/hole valley and layer index, while $E^\mu_{\textbf{Q}} = E^c_{\zeta^\mu L^\mu_e}-E^v_{\zeta^\mu L^\mu_h} + E^\mu_{bind} +  E^\mu_{ \textbf{Q},kin}$ are the excitonic energies, where $E^\mu_{bind}$ are obtained by solving the Wannier equation for a decoupled bilayer system \cite{Ovesen19comphys, Brem20nanoscale}.
Here, $E^{c/v}_{\zeta^\mu L^\mu_e}$ are the conduction and valence band energy, where the bandgap has been fixed to match experimental values, and all the relative valley maxima and minima band alignments have been extracted from DFT calculations \cite{Hagel21prr}.
In addition $E^\mu_{\textbf{Q},kin} = \hbar^2 \textbf{Q}^2/ (2 M^\mu)$ denotes the kinetic energy of the exciton with mass $M^\mu =(m^\mu_e+m^\mu_h)$.
The excitonic tunneling ($\Tb_{\mu \nu}$) between the TMD monolayers can be expressed starting from electronic tunnelling matrix elements
\begin{equation}\label{eq:5}
   \Tb_{\mu \nu} = (\delta_{L^{\mu}_h L^{\nu}_h} (1- \delta_{L^{\mu}_e L^{\nu}_e})  \delta_{\zeta^\mu \zeta^\nu} T^c_{\mu_e,\nu_e} - \delta_{L^{\mu}_e L^{\nu}_e} (1- \delta_{L^{\mu}_h L^{\nu}_h}) \delta_{\zeta^\mu \zeta^\nu} T^v_{\mu_h,\nu_h})  \sum_{\textbf{k}}{ \psi^{\mu*}(\textbf{k} )\psi^{\nu}( \textbf{k}) },
\end{equation}
where $\psi^{\mu}$ is the excitonic wave function of the state $\mu$ defined over the relative momentum between electron and hole.
Furthermore,  $T^\lambda_{i j} = \bra{\lambda i \textbf{p} } H \ket{\lambda j \textbf{p}} (1-\delta_{L_i L_j}) \delta_{\zeta_i \zeta_j}$ are the electronic tunneling elements obtained from DFT calculations \cite{Hagel21prr}. 
Diagonalizing Eq. \ref{eq:4} leads to a new set of hybrid excitonic energies $\Ea^\eta_{\textbf{Q}}$, which are obtained by solving the hybrid eigenvalue equation \cite{Brem20nanoscale, Hagel21prr},
\begin{equation}\label{eq:6}
    E^\mu_{\textbf{Q}} \bm{c}^\eta_\mu(\textbf{Q}) + \sum_{\nu}{ \Tb^{}_{\mu \nu} \bm{c}^\eta_\nu( \textbf{Q}) } = \Ea^\eta_{ \textbf{Q}} \bm{c}^\eta_\mu(\textbf{Q}).
\end{equation}
The diagonalized hybrid exciton Hamiltonian reads \cite{Schmitt22nat, Meneghini22naturalsciences}
\begin{equation}\label{eq:7}
     H= \sum_{\eta}{\Ea^\eta_{\textbf{Q}}  Y^{\eta \dagger}_{\textbf{Q}}  Y^\eta_{\textbf{Q}} } 
\end{equation} 
with the hybrid exciton annihilation/creation operators $Y^{\eta(\dagger)}_{\textbf{Q}} = \sum_{ \mu }{ \bm{c}^\eta_\mu( \textbf{Q}) X^{\mu(\dagger)}_{\textbf{Q}}}$.
Using Eq. \ref{eq:7} we have access to the hybrid exciton energy landscape for the investigated MoS$_2$ homobilayer.

The hybrid exciton-phonon scattering plays a crucial role at the low excitation regime \cite{Brem18scirep,meneghini2022ultrafast}.
The corresponding Hamiltonian can be written as \cite{Brem20nanoscale}
\begin{equation}\label{eq:9}
    H_{Y-ph} = \sum_{j,\textbf{Q},\textbf{q}, \eta,\xi}{ \tilde{\Da}^{\xi\eta}_{j,\textbf{q},\textbf{Q}} Y^{\xi\dagger }_{\textbf{Q + q}}Y^{\eta}_{\textbf{Q}} b_{j,\textbf{q}} } + h.c.
\end{equation}
with the hybrid exciton-phonon coupling $\tilde{\Da}^{\xi\eta}_{j,\textbf{q},\textbf{Q}}$.
The electron-phonon matrix elements, single-particle energies and effective masses are taken from DFPT calculations \cite{PhysRevB.90.045422}.

The excitation of the system through a laser pulse is described semi-classically via the minimal-coupling Hamiltonian that can be written as \cite{Brem20nanoscale}
\begin{equation}
    H_{Y-l} = \sum_{\sigma,\textbf{Q},\eta}{ \textbf{A} \cdot \tilde{\mathcal{M}}^\eta_{\sigma \textbf{Q}} Y^\eta_{\textbf{Q}_{\parallel} } } + h.c.
\end{equation}
with zjr hybrid exciton-light coupling $\tilde{\mathcal{M}}^\eta_{\sigma \textbf{Q}}$.
Details on the transformation and the definition of the hybrid interaction matrix elements and couplings are given in Ref. \cite{Brem20nanoscale, Hagel21prr}.

The dynamics of the system is studied by solving the Heisenberg equation of motion for the hybrid occupation $N^\eta = \langle Y^{\eta\dagger }Y^{\eta}  \rangle$, using the full hybrid exciton Hamiltonian $H = H_Y + H_{Y-ph}+ H_{Y-l}$, and truncating the Martin-Schwinger hierarchy using a second order Born-Markov approximation \cite{kira2006many,haug2009quantum,malic2013graphene}.
We separate the coherent $P^{\eta}_{\textbf{Q}} =\langle Y^{\eta \dagger}_{ \textbf{Q}}\rangle$ and the incoherent hybrid populations $\delta N^{\eta }_{\textbf{Q}} = \langle Y^{\eta\dagger }_{\textbf{Q}}Y^{\eta}_{\textbf{Q}}\rangle-\langle Y^{\eta \dagger}_{\textbf{Q}}\rangle \langle Y^{\eta}_{\textbf{Q}} \rangle = N^{\eta }_{\textbf{Q}} - |P^{\eta}_{\textbf{Q}}|^2$ leading to the following semiconductor Bloch equations
\begin{align}\label{eq:10}
\begin{split}
    i\hbar \partial_t P^\eta_0 &= -(\Ea^\eta_0 + i \Gamma^\eta_0)P^\eta_0 -  \tilde{\mathcal{M}}^\eta_0 \cdot \textbf{A}(t)\\[6pt]
    \delta \dot{N}^\eta_{\textbf{Q}} &= \sum_{\xi}{ W^{\xi\eta}_{\textbf{0 Q}}}  \abs{P^{\eta}_\textbf{0}}^2  + \sum_{\xi, {\textbf{Q}'}}{ \left( W^{\xi\eta}_{\textbf{Q}' \textbf{Q}}  \delta N^\xi_{\textbf{Q}'} - W^{\eta\xi}_{{\textbf{Q Q}' }} \delta N^\eta_{\textbf{Q}} \right) }
\end{split}
\end{align}
where the appearing scattering rates are related via $2\Gamma^\eta_{\textbf{Q}}/\hbar =  \sum_{\textbf{Q}'\xi}{ W^{\xi\eta}_{\textbf{Q} \textbf{Q}'}}$ with the phonon mediated scattering tensor defined as 
$$W^{\eta\xi}_{{\textbf{Q Q}'}} = \frac{2\pi}{\hbar}
\sum_{j,\pm}|\tilde{\Da}^{\eta\xi}_{j,\textbf{Q}'-\textbf{Q}} |^2 \left( \frac{1}{2} \pm
\frac{1}{2} + n^{ph}_{j,\textbf{Q}'-\textbf{Q}} \right) \delta \left( \Ea^{\xi}_{\textbf{Q}'} -
\Ea^\eta_\textbf{Q} \mp   \hbar\Omega_{j \textbf{Q}'-\textbf{Q}} \right).$$

To model the ARPES signal and have a direct access to the momentum and energy map, we apply the Fermis-Golden rule yielding
\begin{align}\label{eq:arpes1}
    \mathcal{I}(\textbf{k},h\nu; t) \propto \sum_{i,f} \lvert \bra{f \textbf{k}} H_{int} \ket{i}\rvert^2 N_i(t)
    \delta \left(E_{f \textbf{k}}- E_i - h\nu \right)
\end{align}
where $\ket{i/f}$ are the initial/final states of the system where we consider eigenstates of the two-body Hamiltonian with initial/final eigenergies $E_{i/f}$.
The energy of the photon is denoted by $h\nu$ and the initial state occupation by $N_i(t)$.
The initial states in our system are hybrid excitons, i.e. $\ket{i}= \ket{\eta \textbf{Q}}$ and $N_i(t) = N^\eta_{\textbf{Q}}$, while the final state consists of an ejected free electron and a left-behind hole in a hybridized valence band of the bilayer.
To evaluate the expectation value in Eq.\ref{eq:arpes1} we express the initial and final states in terms of electronic creation and annihilation ($a/a^\dagger$) operators.

The final state can be expressed as the product of a free electron state $\ket{\textbf{k}}$ and a hybrid hole state $\ket{\gamma \lambda \textbf{p} }$ yielding 
\begin{equation}
    \ket{f} = \ket{{\textbf{k}},\gamma v{\textbf{p}} } = \sum_l g^\gamma_{l \textbf{p}} a^\dagger_\textbf{k} a^{ }_{v \textbf{p} l \xi_h} \ket{0}
\end{equation} 
where the hybrid hole state is obtained by solving the eigenvalue problem for the electronic bilayer Hamiltonian
\begin{equation}
    H = \sum_{\textbf{k} l \lambda} \mathcal{E}^{ }_{\lambda l \textbf{k}} a^\dagger_{\lambda\textbf{k} l } a^{ }_{\lambda\textbf{k} l} +   \sum_{\textbf{k} l l' \lambda} \mathcal{T}^{ }_{\lambda l l^\prime} a^\dagger_{\lambda\textbf{k} l^\prime}a^{ }_{\lambda\textbf{k} l}
\end{equation}
with $\lambda = c,v$ indicating the band index and $a^{(\dagger)}_{\lambda\textbf{k} l }$ conduction/valence band annihilation(creation) operators and $\mathcal{T}^{ }_{\lambda l l'}$ band-dependent tunneling strength between the two layers $l l'$.
The solution of the eigenvalue problem leads us to a set of hybridized valence and conduction bands
\begin{equation}
    E^\lambda_{\textbf{k}\gamma} =   \frac{1}{2}   \left(\mathcal{E}_{\lambda, 1 \textbf{k}} +\mathcal{E}_{\lambda, 2\textbf{k}}\right) \pm   \frac{1}{2}\sqrt{\left(\mathcal{E}_{\lambda, 1 \textbf{k}} -\mathcal{E}_{\lambda, 2 \textbf{k}}\right)^2 + 4 {|\mathcal{T}_{\lambda, 12}|}^2 } 
\end{equation}
with $\gamma = (\pm,\xi)$ labeling the two new states ($\pm$ solutions) and the valley index $\xi$.
The corresponding eigenvectors are obtained from the same 2x2 eigenvalue problem which we write as a superposition of the old monolayer states as
\begin{equation*}
    \ket{\gamma \lambda \textbf{p} } = \sum_{l} g^\gamma_{l \textbf{p}} a^\dagger_{\lambda  \textbf{p} l \xi} \ket{0}
\end{equation*}
with the mixing coefficients $g^\gamma_{l}$.

The initial state can be expressed starting from the hybrid exciton state $\ket{\eta \textbf{Q}}$ and perform a series of backward transformation from hybrid exciton operators to exciton operators arriving finally to electron operators $Y^\dagger \rightarrow X^\dagger \rightarrow a^\dagger a$, obtaining
\begin{align*}
    \ket{i} = \ket{\eta \textbf{Q}} &= Y^\dagger_{\eta \textbf{Q}} \ket{0} = \sum_{\mu} {\bm{c}^{\eta *}_{\mu}}(\textbf{Q}) X^\dagger_{\mu\textbf{Q}}\ket{0} = \sum_{\mu\textbf{k}} {\bm{c}^{\eta *}_{\mu}}(\textbf{Q}) {\psi^{\mu *}}(\textbf{k}) a^\dagger_{c,\textbf{k}+\tilde{m}_e \textbf{Q}, \mu_e} a^{ }_{v,\textbf{k}-\tilde{m}_h \textbf{Q}, \mu_h} \ket{0}
\end{align*}
with the compound index $\eta = (n,\xi)$ describing the hybrid degrees of freedom, ${\bm{c}^{\eta *}_{\mu}}(\textbf{Q})$ denoting excitonic mixing coefficients, $\mu = (L,\xi)$ describing the excitonic degrees of freedom with $L = (l_e,l_h)$ and $\xi = (\xi_e,\xi_h)$.
We use the notation $\mu_{e/h}$ to refer the quantum numbers inside $\mu$ labeled by $e/h$.
Note that we include  only the lowest 1s excitonic states.

Inserting the initial and final states in Eq. (\ref{eq:arpes1}) we obtain
\begin{align}
    \mathcal{I}({\textbf{k}},h\nu) \propto \sum_{\substack{\eta\gamma \\{\textbf{p}}\textbf{Q}}} \lvert \bra{{\textbf{k}}, \gamma v {\textbf{p}} } H_{int} \ket{\eta \textbf{Q}}\rvert^2 \cdot N_{\textbf{Q}}^\eta(t) \text{ } \delta\left(E^{e}_\textbf{k} - E^{v}_{\gamma,\textbf{p} } - E^{X}_{\eta,\textbf{Q}} - h\nu\right)
\end{align}
where $\textbf{p}$ is the hole momentum.
Furthermore, $N^\eta_{\textbf{Q}}(t)$ denotes the hybrid exciton time-dependent occupation for the hX state $\eta$ at the center-of-mass momentum $\textbf{Q}$.
Moreover, $E^{e}_\textbf{k}$ corresponds to the free electron energy, $E^{v}_{\gamma,\textbf{p}}$ to the hybrid valence band energy, and  $E^{X}_{\eta,\textbf{Q}}$ to the hybrid exciton energy.
The electron-light Hamiltonian reads
\begin{align}
    H_{int} = \sum_{\textbf{p} \textbf{k} \gamma} \mathcal{M}^{}_{\textbf{p}\textbf{k}\xi_e} a^\dagger_{\mathrm{f}\textbf{p}} a^{\gamma}_{c\textbf{k}} =\sum_{\textbf{p} \textbf{k} \gamma l} g^{\gamma}_{l \textbf{k}} \mathcal{M}^{}_{\textbf{p}\textbf{k}\xi_e} a^\dagger_{ \textbf{p}} a^{ }_{c \textbf{k} l \xi_e}
\end{align}
with the optical matrix element $\mathcal{M}_{\textbf{p}\textbf{k}\xi_e}$ containing the optical selection rules.
By evaluating $ \bra{ \textbf{k}, \gamma v {\textbf{p}}} H_{int} \ket{\eta \textbf{Q}}$ with the above definition of initial and final states,
we can rewrite the total ARPES signal in the following form
\begin{equation}\label{eq:inten_arpes}
    \mathcal{I}({\textbf{k}},h\nu;t) \propto \sum_{\substack{\eta,\gamma,{\textbf{p}} }} \lvert  \mathcal{G}^{\eta \gamma}_{\textbf{p} \tilde{\textbf{k}}}\rvert^2 N^\eta_{\tilde{\textbf{k}} -\textbf{p}}(t) \text{ } \delta\left(E^{e}_\textbf{k} - E^{v}_{\gamma,\textbf{p}} - E^{X}_{\eta,\tilde{\textbf{k}} -\textbf{p}} - h\nu\right)
\end{equation}%
with
\begin{align}\label{eq:temp}
 \mathcal{G}^{\eta\gamma }_{\textbf{p} \tilde{\textbf{k}}} =  \sum_\mu g^{\gamma *}_{l^\mu_h \textbf{p}}  \tilde{\mathcal{M}} {\bm{c}^{\eta *}_{\mu}}(\tilde{\textbf{k}} -\textbf{p}) {\psi^{\mu *}}(\tilde{m}_e\textbf{p} + \tilde{m}_h\tilde{\textbf{k}} ) \cdot \delta_{\xi^\gamma_h,\xi^\mu_h}.
\end{align}
where we used $\tilde{\textbf{k}} = \textbf{k} - \xi_e$, and due to conservation of the total electron momentum we neglect all other momentum dependencies resulting in $\mathcal{M}_{\textbf{p}\textbf{k}\xi} = \tilde{\mathcal{M}} \delta_{\textbf{p}_{\parallel},\textbf{k}+\xi}$.
The new coupling $\mathcal{G}^{\eta\gamma }_{\textbf{p} \tilde{\textbf{k}}}$ contains the momentum dependence of the ARPES signal, i.e. the superposition of excitonic wave functions weighted by  mixing coefficients and hole hybridization coefficients.
Note that ARPES signals stemming from different electron valleys are additionally weighted by different photoemission matrix elements $\mathcal{M}$, which is neglected here.
We refer to \cite{Meneghini23acspho} for a complete derivation and definition of the aforementioned quantities.

\section{ACKNOWLEDGEMENTS}

This work was funded by the Deutsche Forschungsgemeinschaft (DFG, German Research Foundation) - 432680300/SFB 1456 (project B01), 217133147/SFB 1073 (projects B07 and B10), 223848855/SFB 1083 (project B9), the Priority Program SPP 2244 “2DMP” (Project No. 535247173), and the project 542873258. A.A. and S.H. acknowledge funding from EPSRC (EP/T001038/1, EP/P005152/1). A.A. acknowledges financial support by the Saudi Arabian Ministry of Higher Education. K.W. and T.T. acknowledge support from the JSPS KAKENHI (Grant Numbers 21H05233 and 23H02052), the CREST (JPMJCR24A5), JST and World Premier International Research Center Initiative (WPI), MEXT, Japan.

\section{DATA AVAILABILITY}
The experimental data and evaluation will be published through GRO.Data.

\section{AUTHOR CONTRIBUTIONS}
J.L., D.St., S.Ho, R.T.W., E.M, G.S.M.J., S.M., and M.R. conceived the research.
P.W., A.A., and A.S. fabricated the sample.
P.W., W.B., J.P.B., D.S. and M.M. carried out the time-resolved momentum microscopy experiments. 
P.W. analyzed the experimental data and prepared the figures.
G.M. carried out the theoretical calculations guided by E.M..
All authors discussed the results. 
M.R. and S.M. were responsible for the overall project direction. 
S.M. and M.R. wrote the manuscript with contributions from all co-authors. 
K.W. and T.T. synthesized the hBN crystals.

\setcounter{figure}{0}
\captionsetup[figure]{name={Extended Fig.}, font={small, stretch=1.2}, labelfont={bf}}

\begin{figure}[h]
    \begin{center}
	    \includegraphics{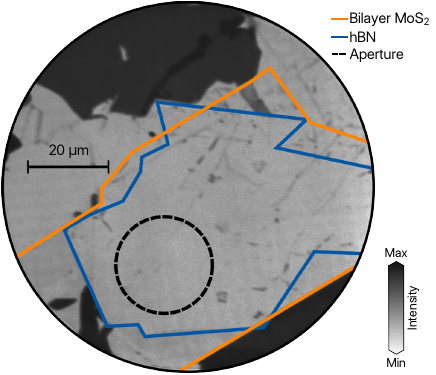}
    \end{center}
    \caption{
	Real space photoemission electron microscopy image of the 2H-MoS$_2$ homobilayer sample. 
	Black dashed circle marks the aperture position used for spatially selective momentum microscopy measurements. 
	Orange outline marks the bilayer 2H-MoS$_2$ flake, the blue outline marks the hBN buffer layer. Black, high intensity regions stem from the Nb-STO substrate.
    }
\label{fig:sample_img}
\end{figure}

\begin{figure}[h]
  \begin{center}
    \includegraphics{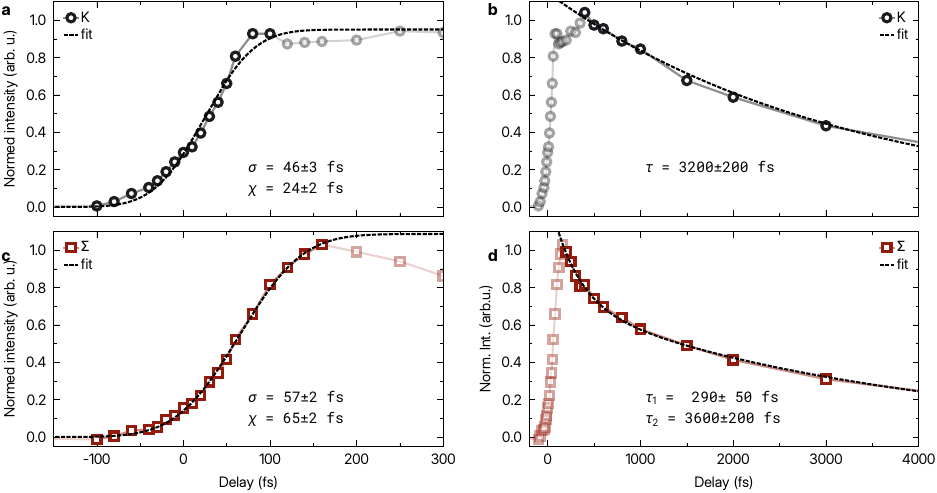}
  \end{center}
  \caption{
  \textbf{Fit of rise and decay of excited state signal}.
  a) Rise of the exited state signal captured at the K valleys
  b) Single exponential fit of the decay of the K valley signal.
  c) Same as a) for the signal at the $\Sigma$ valley.
  d) Double exponential fit of the decay of the $\Sigma$ valley signal.
  }\label{fig:time_fits}
\end{figure}

\begin{figure}[h]
    \begin{center}
        \includegraphics{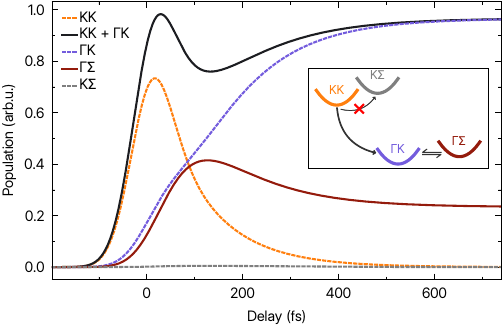}
    \end{center}
	\caption{Time-resolved population of selected exciton states obtained from microscopic modelling without applying normalization.}
	\label{fig:theory_time_traces_no_norm}
\end{figure}


\begin{thebibliography}{66}%
\makeatletter
\providecommand \@ifxundefined [1]{%
 \@ifx{#1\undefined}
}%
\providecommand \@ifnum [1]{%
 \ifnum #1\expandafter \@firstoftwo
 \else \expandafter \@secondoftwo
 \fi
}%
\providecommand \@ifx [1]{%
 \ifx #1\expandafter \@firstoftwo
 \else \expandafter \@secondoftwo
 \fi
}%
\providecommand \natexlab [1]{#1}%
\providecommand \enquote  [1]{``#1''}%
\providecommand \bibnamefont  [1]{#1}%
\providecommand \bibfnamefont [1]{#1}%
\providecommand \citenamefont [1]{#1}%
\providecommand \href@noop [0]{\@secondoftwo}%
\providecommand \href [0]{\begingroup \@sanitize@url \@href}%
\providecommand \@href[1]{\@@startlink{#1}\@@href}%
\providecommand \@@href[1]{\endgroup#1\@@endlink}%
\providecommand \@sanitize@url [0]{\catcode `\\12\catcode `\$12\catcode
  `\&12\catcode `\#12\catcode `\^12\catcode `\_12\catcode `\%12\relax}%
\providecommand \@@startlink[1]{}%
\providecommand \@@endlink[0]{}%
\providecommand \url  [0]{\begingroup\@sanitize@url \@url }%
\providecommand \@url [1]{\endgroup\@href {#1}{\urlprefix }}%
\providecommand \urlprefix  [0]{URL }%
\providecommand \Eprint [0]{\href }%
\providecommand \doibase [0]{https://doi.org/}%
\providecommand \selectlanguage [0]{\@gobble}%
\providecommand \bibinfo  [0]{\@secondoftwo}%
\providecommand \bibfield  [0]{\@secondoftwo}%
\providecommand \translation [1]{[#1]}%
\providecommand \BibitemOpen [0]{}%
\providecommand \bibitemStop [0]{}%
\providecommand \bibitemNoStop [0]{.\EOS\space}%
\providecommand \EOS [0]{\spacefactor3000\relax}%
\providecommand \BibitemShut  [1]{\csname bibitem#1\endcsname}%
\let\auto@bib@innerbib\@empty
%</preamble>
\bibitem [{\citenamefont {Liang}\ \emph {et~al.}(2020)\citenamefont {Liang},
  \citenamefont {Cheng}, \citenamefont {Cui},\ and\ \citenamefont
  {Miao}}]{Liang20advmat}%
  \BibitemOpen
  \bibfield  {author} {\bibinfo {author} {\bibfnamefont {S.-J.}\ \bibnamefont
  {Liang}}, \bibinfo {author} {\bibfnamefont {B.}~\bibnamefont {Cheng}},
  \bibinfo {author} {\bibfnamefont {X.}~\bibnamefont {Cui}},\ and\ \bibinfo
  {author} {\bibfnamefont {F.}~\bibnamefont {Miao}},\ }\bibfield  {title}
  {\bibinfo {title} {Van der {Waals} heterostructures for high-performance
  device applications: Challenges and opportunities},\ }\href
  {https://doi.org/https://doi.org/10.1002/adma.201903800} {\bibfield
  {journal} {\bibinfo  {journal} {Advanced Materials}\ }\textbf {\bibinfo
  {volume} {32}},\ \bibinfo {pages} {1903800} (\bibinfo {year}
  {2020})}\BibitemShut {NoStop}%
\bibitem [{\citenamefont {Mueller}\ and\ \citenamefont
  {Malic}(2018)}]{Mueller2018npj}%
  \BibitemOpen
  \bibfield  {author} {\bibinfo {author} {\bibfnamefont {T.}~\bibnamefont
  {Mueller}}\ and\ \bibinfo {author} {\bibfnamefont {E.}~\bibnamefont
  {Malic}},\ }\bibfield  {title} {\bibinfo {title} {Exciton physics and device
  application of two-dimensional transition metal dichalcogenide
  semiconductors},\ }\href {https://doi.org/10.1038/s41699-018-0074-2}
  {\bibfield  {journal} {\bibinfo  {journal} {npj 2D Materials and
  Applications}\ }\textbf {\bibinfo {volume} {2}},\ \bibinfo {pages} {29}
  (\bibinfo {year} {2018})}\BibitemShut {NoStop}%
\bibitem [{\citenamefont {Wang}\ \emph {et~al.}(2018)\citenamefont {Wang},
  \citenamefont {Chernikov}, \citenamefont {Glazov}, \citenamefont {Heinz},
  \citenamefont {Marie}, \citenamefont {Amand},\ and\ \citenamefont
  {Urbaszek}}]{Wang18rmp}%
  \BibitemOpen
  \bibfield  {author} {\bibinfo {author} {\bibfnamefont {G.}~\bibnamefont
  {Wang}}, \bibinfo {author} {\bibfnamefont {A.}~\bibnamefont {Chernikov}},
  \bibinfo {author} {\bibfnamefont {M.~M.}\ \bibnamefont {Glazov}}, \bibinfo
  {author} {\bibfnamefont {T.~F.}\ \bibnamefont {Heinz}}, \bibinfo {author}
  {\bibfnamefont {X.}~\bibnamefont {Marie}}, \bibinfo {author} {\bibfnamefont
  {T.}~\bibnamefont {Amand}},\ and\ \bibinfo {author} {\bibfnamefont
  {B.}~\bibnamefont {Urbaszek}},\ }\bibfield  {title} {\bibinfo {title}
  {{Colloquium: Excitons in atomically thin transition metal
  dichalcogenides}},\ }\href {https://doi.org/10.1103/RevModPhys.90.021001}
  {\bibfield  {journal} {\bibinfo  {journal} {Reviews of Modern Physics}\
  }\textbf {\bibinfo {volume} {90}},\ \bibinfo {pages} {021001} (\bibinfo
  {year} {2018})}\BibitemShut {NoStop}%
\bibitem [{\citenamefont {Perea-Causin}\ \emph {et~al.}(2022)\citenamefont
  {Perea-Causin}, \citenamefont {Erkensten}, \citenamefont {Fitzgerald},
  \citenamefont {Thompson}, \citenamefont {Rosati}, \citenamefont {Brem},\ and\
  \citenamefont {Malic}}]{Perea22apl}%
  \BibitemOpen
  \bibfield  {author} {\bibinfo {author} {\bibfnamefont {R.}~\bibnamefont
  {Perea-Causin}}, \bibinfo {author} {\bibfnamefont {D.}~\bibnamefont
  {Erkensten}}, \bibinfo {author} {\bibfnamefont {J.~M.}\ \bibnamefont
  {Fitzgerald}}, \bibinfo {author} {\bibfnamefont {J.~J.~P.}\ \bibnamefont
  {Thompson}}, \bibinfo {author} {\bibfnamefont {R.}~\bibnamefont {Rosati}},
  \bibinfo {author} {\bibfnamefont {S.}~\bibnamefont {Brem}},\ and\ \bibinfo
  {author} {\bibfnamefont {E.}~\bibnamefont {Malic}},\ }\bibfield  {title}
  {\bibinfo {title} {Exciton optics, dynamics, and transport in atomically thin
  semiconductors},\ }\href {https://doi.org/10.1063/5.0107665} {\bibfield
  {journal} {\bibinfo  {journal} {APL Materials}\ }\textbf {\bibinfo {volume}
  {10}},\ \bibinfo {pages} {100701} (\bibinfo {year} {2022})}\BibitemShut
  {NoStop}%
\bibitem [{\citenamefont {Ruiz-Tijerina}\ and\ \citenamefont
  {Fal'ko}(2019)}]{Ruiz19prb}%
  \BibitemOpen
  \bibfield  {author} {\bibinfo {author} {\bibfnamefont {D.~A.}\ \bibnamefont
  {Ruiz-Tijerina}}\ and\ \bibinfo {author} {\bibfnamefont {V.~I.}\ \bibnamefont
  {Fal'ko}},\ }\bibfield  {title} {\bibinfo {title} {Interlayer hybridization
  and moir\'e superlattice minibands for electrons and excitons in
  heterobilayers of transition-metal dichalcogenides},\ }\href
  {https://doi.org/10.1103/PhysRevB.99.125424} {\bibfield  {journal} {\bibinfo
  {journal} {Physical Review B}\ }\textbf {\bibinfo {volume} {99}},\ \bibinfo
  {pages} {125424} (\bibinfo {year} {2019})}\BibitemShut {NoStop}%
\bibitem [{\citenamefont {Wang}\ \emph {et~al.}(2017)\citenamefont {Wang},
  \citenamefont {Wang}, \citenamefont {Yao}, \citenamefont {Liu},\ and\
  \citenamefont {Yu}}]{Wang17prb}%
  \BibitemOpen
  \bibfield  {author} {\bibinfo {author} {\bibfnamefont {Y.}~\bibnamefont
  {Wang}}, \bibinfo {author} {\bibfnamefont {Z.}~\bibnamefont {Wang}}, \bibinfo
  {author} {\bibfnamefont {W.}~\bibnamefont {Yao}}, \bibinfo {author}
  {\bibfnamefont {G.-B.}\ \bibnamefont {Liu}},\ and\ \bibinfo {author}
  {\bibfnamefont {H.}~\bibnamefont {Yu}},\ }\bibfield  {title} {\bibinfo
  {title} {{Interlayer coupling in commensurate and incommensurate bilayer
  structures of transition-metal dichalcogenides}},\ }\href
  {https://doi.org/10.1103/PhysRevB.95.115429} {\bibfield  {journal} {\bibinfo
  {journal} {Physical Review B}\ }\textbf {\bibinfo {volume} {95}},\ \bibinfo
  {pages} {115429} (\bibinfo {year} {2017})}\BibitemShut {NoStop}%
\bibitem [{\citenamefont {Chernikov}\ \emph {et~al.}(2015)\citenamefont
  {Chernikov}, \citenamefont {van~der Zande}, \citenamefont {Hill},
  \citenamefont {Rigosi}, \citenamefont {Velauthapillai}, \citenamefont
  {Hone},\ and\ \citenamefont {Heinz}}]{Chernikov15prl}%
  \BibitemOpen
  \bibfield  {author} {\bibinfo {author} {\bibfnamefont {A.}~\bibnamefont
  {Chernikov}}, \bibinfo {author} {\bibfnamefont {A.~M.}\ \bibnamefont {van~der
  Zande}}, \bibinfo {author} {\bibfnamefont {H.~M.}\ \bibnamefont {Hill}},
  \bibinfo {author} {\bibfnamefont {A.~F.}\ \bibnamefont {Rigosi}}, \bibinfo
  {author} {\bibfnamefont {A.}~\bibnamefont {Velauthapillai}}, \bibinfo
  {author} {\bibfnamefont {J.}~\bibnamefont {Hone}},\ and\ \bibinfo {author}
  {\bibfnamefont {T.~F.}\ \bibnamefont {Heinz}},\ }\bibfield  {title} {\bibinfo
  {title} {Electrical tuning of exciton binding energies in monolayer
  {WS}$_2$},\ }\href {https://doi.org/10.1103/PhysRevLett.115.126802}
  {\bibfield  {journal} {\bibinfo  {journal} {Physical Review Letters}\
  }\textbf {\bibinfo {volume} {115}},\ \bibinfo {pages} {126802} (\bibinfo
  {year} {2015})}\BibitemShut {NoStop}%
\bibitem [{\citenamefont {Raja}\ \emph {et~al.}(2017)\citenamefont {Raja},
  \citenamefont {Chaves}, \citenamefont {Yu}, \citenamefont {Arefe},
  \citenamefont {Hill}, \citenamefont {Rigosi}, \citenamefont {Berkelbach},
  \citenamefont {Nagler}, \citenamefont {Schüller}, \citenamefont {Korn},
  \citenamefont {Nuckolls}, \citenamefont {Hone}, \citenamefont {Brus},
  \citenamefont {Heinz}, \citenamefont {Reichman},\ and\ \citenamefont
  {Chernikov}}]{Raja17natcomm}%
  \BibitemOpen
  \bibfield  {author} {\bibinfo {author} {\bibfnamefont {A.}~\bibnamefont
  {Raja}}, \bibinfo {author} {\bibfnamefont {A.}~\bibnamefont {Chaves}},
  \bibinfo {author} {\bibfnamefont {J.}~\bibnamefont {Yu}}, \bibinfo {author}
  {\bibfnamefont {G.}~\bibnamefont {Arefe}}, \bibinfo {author} {\bibfnamefont
  {H.~M.}\ \bibnamefont {Hill}}, \bibinfo {author} {\bibfnamefont {A.~F.}\
  \bibnamefont {Rigosi}}, \bibinfo {author} {\bibfnamefont {T.~C.}\
  \bibnamefont {Berkelbach}}, \bibinfo {author} {\bibfnamefont
  {P.}~\bibnamefont {Nagler}}, \bibinfo {author} {\bibfnamefont
  {C.}~\bibnamefont {Schüller}}, \bibinfo {author} {\bibfnamefont
  {T.}~\bibnamefont {Korn}}, \bibinfo {author} {\bibfnamefont {C.}~\bibnamefont
  {Nuckolls}}, \bibinfo {author} {\bibfnamefont {J.}~\bibnamefont {Hone}},
  \bibinfo {author} {\bibfnamefont {L.~E.}\ \bibnamefont {Brus}}, \bibinfo
  {author} {\bibfnamefont {T.~F.}\ \bibnamefont {Heinz}}, \bibinfo {author}
  {\bibfnamefont {D.~R.}\ \bibnamefont {Reichman}},\ and\ \bibinfo {author}
  {\bibfnamefont {A.}~\bibnamefont {Chernikov}},\ }\bibfield  {title} {\bibinfo
  {title} {Coulomb engineering of the bandgap and excitons in two-dimensional
  materials},\ }\href {https://doi.org/10.1038/ncomms15251} {\bibfield
  {journal} {\bibinfo  {journal} {Nature Communications}\ }\textbf {\bibinfo
  {volume} {8}},\ \bibinfo {pages} {15251} (\bibinfo {year}
  {2017})}\BibitemShut {NoStop}%
\bibitem [{\citenamefont {Alexeev}\ \emph {et~al.}(2019)\citenamefont
  {Alexeev}, \citenamefont {Ruiz-Tijerina}, \citenamefont {Danovich},
  \citenamefont {Hamer}, \citenamefont {Terry}, \citenamefont {Nayak},
  \citenamefont {Ahn}, \citenamefont {Pak}, \citenamefont {Lee}, \citenamefont
  {Sohn}, \citenamefont {Molas}, \citenamefont {Koperski}, \citenamefont
  {Watanabe}, \citenamefont {Taniguchi}, \citenamefont {Novoselov},
  \citenamefont {Gorbachev}, \citenamefont {Shin}, \citenamefont {Fal’ko},\
  and\ \citenamefont {Tartakovskii}}]{Alexeev19nat}%
  \BibitemOpen
  \bibfield  {author} {\bibinfo {author} {\bibfnamefont {E.~M.}\ \bibnamefont
  {Alexeev}}, \bibinfo {author} {\bibfnamefont {D.~A.}\ \bibnamefont
  {Ruiz-Tijerina}}, \bibinfo {author} {\bibfnamefont {M.}~\bibnamefont
  {Danovich}}, \bibinfo {author} {\bibfnamefont {M.~J.}\ \bibnamefont {Hamer}},
  \bibinfo {author} {\bibfnamefont {D.~J.}\ \bibnamefont {Terry}}, \bibinfo
  {author} {\bibfnamefont {P.~K.}\ \bibnamefont {Nayak}}, \bibinfo {author}
  {\bibfnamefont {S.}~\bibnamefont {Ahn}}, \bibinfo {author} {\bibfnamefont
  {S.}~\bibnamefont {Pak}}, \bibinfo {author} {\bibfnamefont {J.}~\bibnamefont
  {Lee}}, \bibinfo {author} {\bibfnamefont {J.~I.}\ \bibnamefont {Sohn}},
  \bibinfo {author} {\bibfnamefont {M.~R.}\ \bibnamefont {Molas}}, \bibinfo
  {author} {\bibfnamefont {M.}~\bibnamefont {Koperski}}, \bibinfo {author}
  {\bibfnamefont {K.}~\bibnamefont {Watanabe}}, \bibinfo {author}
  {\bibfnamefont {T.}~\bibnamefont {Taniguchi}}, \bibinfo {author}
  {\bibfnamefont {K.~S.}\ \bibnamefont {Novoselov}}, \bibinfo {author}
  {\bibfnamefont {R.~V.}\ \bibnamefont {Gorbachev}}, \bibinfo {author}
  {\bibfnamefont {H.~S.}\ \bibnamefont {Shin}}, \bibinfo {author}
  {\bibfnamefont {V.~I.}\ \bibnamefont {Fal’ko}},\ and\ \bibinfo {author}
  {\bibfnamefont {A.~I.}\ \bibnamefont {Tartakovskii}},\ }\bibfield  {title}
  {\bibinfo {title} {{Resonantly hybridized excitons in moiré superlattices in
  van der {Waals} heterostructures}},\ }\href
  {https://doi.org/10.1038/s41586-019-0986-9} {\bibfield  {journal} {\bibinfo
  {journal} {Nature}\ }\textbf {\bibinfo {volume} {567}},\ \bibinfo {pages}
  {81} (\bibinfo {year} {2019})}\BibitemShut {NoStop}%
\bibitem [{\citenamefont {Tran}\ \emph {et~al.}(2019)\citenamefont {Tran},
  \citenamefont {Moody}, \citenamefont {Wu}, \citenamefont {Lu}, \citenamefont
  {Choi}, \citenamefont {Kim}, \citenamefont {Rai}, \citenamefont {Sanchez},
  \citenamefont {Quan}, \citenamefont {Singh}, \citenamefont {Embley},
  \citenamefont {Zepeda}, \citenamefont {Campbell}, \citenamefont {Autry},
  \citenamefont {Taniguchi}, \citenamefont {Watanabe}, \citenamefont {Lu},
  \citenamefont {Banerjee}, \citenamefont {Silverman}, \citenamefont {Kim},
  \citenamefont {Tutuc}, \citenamefont {Yang}, \citenamefont {MacDonald},\ and\
  \citenamefont {Li}}]{Tran19nat}%
  \BibitemOpen
  \bibfield  {author} {\bibinfo {author} {\bibfnamefont {K.}~\bibnamefont
  {Tran}}, \bibinfo {author} {\bibfnamefont {G.}~\bibnamefont {Moody}},
  \bibinfo {author} {\bibfnamefont {F.}~\bibnamefont {Wu}}, \bibinfo {author}
  {\bibfnamefont {X.}~\bibnamefont {Lu}}, \bibinfo {author} {\bibfnamefont
  {J.}~\bibnamefont {Choi}}, \bibinfo {author} {\bibfnamefont {K.}~\bibnamefont
  {Kim}}, \bibinfo {author} {\bibfnamefont {A.}~\bibnamefont {Rai}}, \bibinfo
  {author} {\bibfnamefont {D.~A.}\ \bibnamefont {Sanchez}}, \bibinfo {author}
  {\bibfnamefont {J.}~\bibnamefont {Quan}}, \bibinfo {author} {\bibfnamefont
  {A.}~\bibnamefont {Singh}}, \bibinfo {author} {\bibfnamefont
  {J.}~\bibnamefont {Embley}}, \bibinfo {author} {\bibfnamefont
  {A.}~\bibnamefont {Zepeda}}, \bibinfo {author} {\bibfnamefont
  {M.}~\bibnamefont {Campbell}}, \bibinfo {author} {\bibfnamefont
  {T.}~\bibnamefont {Autry}}, \bibinfo {author} {\bibfnamefont
  {T.}~\bibnamefont {Taniguchi}}, \bibinfo {author} {\bibfnamefont
  {K.}~\bibnamefont {Watanabe}}, \bibinfo {author} {\bibfnamefont
  {N.}~\bibnamefont {Lu}}, \bibinfo {author} {\bibfnamefont {S.~K.}\
  \bibnamefont {Banerjee}}, \bibinfo {author} {\bibfnamefont {K.~L.}\
  \bibnamefont {Silverman}}, \bibinfo {author} {\bibfnamefont {S.}~\bibnamefont
  {Kim}}, \bibinfo {author} {\bibfnamefont {E.}~\bibnamefont {Tutuc}}, \bibinfo
  {author} {\bibfnamefont {L.}~\bibnamefont {Yang}}, \bibinfo {author}
  {\bibfnamefont {A.~H.}\ \bibnamefont {MacDonald}},\ and\ \bibinfo {author}
  {\bibfnamefont {X.}~\bibnamefont {Li}},\ }\bibfield  {title} {\bibinfo
  {title} {Evidence for moiré excitons in van der {Waals} heterostructures},\
  }\href {https://doi.org/10.1038/s41586-019-0975-z} {\bibfield  {journal}
  {\bibinfo  {journal} {Nature}\ }\textbf {\bibinfo {volume} {567}},\ \bibinfo
  {pages} {71} (\bibinfo {year} {2019})}\BibitemShut {NoStop}%
\bibitem [{\citenamefont {Seyler}\ \emph {et~al.}(2019)\citenamefont {Seyler},
  \citenamefont {Rivera}, \citenamefont {Yu}, \citenamefont {Wilson},
  \citenamefont {Ray}, \citenamefont {Mandrus}, \citenamefont {Yan},
  \citenamefont {Yao},\ and\ \citenamefont {Xu}}]{Seyler19nat}%
  \BibitemOpen
  \bibfield  {author} {\bibinfo {author} {\bibfnamefont {K.~L.}\ \bibnamefont
  {Seyler}}, \bibinfo {author} {\bibfnamefont {P.}~\bibnamefont {Rivera}},
  \bibinfo {author} {\bibfnamefont {H.}~\bibnamefont {Yu}}, \bibinfo {author}
  {\bibfnamefont {N.~P.}\ \bibnamefont {Wilson}}, \bibinfo {author}
  {\bibfnamefont {E.~L.}\ \bibnamefont {Ray}}, \bibinfo {author} {\bibfnamefont
  {D.~G.}\ \bibnamefont {Mandrus}}, \bibinfo {author} {\bibfnamefont
  {J.}~\bibnamefont {Yan}}, \bibinfo {author} {\bibfnamefont {W.}~\bibnamefont
  {Yao}},\ and\ \bibinfo {author} {\bibfnamefont {X.}~\bibnamefont {Xu}},\
  }\bibfield  {title} {\bibinfo {title} {{Signatures of moiré-trapped valley
  excitons in {MoSe}$_2$/{WSe}$_2$ heterobilayers}},\ }\href
  {https://doi.org/10.1038/s41586-019-0957-1} {\bibfield  {journal} {\bibinfo
  {journal} {Nature}\ }\textbf {\bibinfo {volume} {567}},\ \bibinfo {pages}
  {66} (\bibinfo {year} {2019})}\BibitemShut {NoStop}%
\bibitem [{\citenamefont {Xu}\ \emph {et~al.}(2020)\citenamefont {Xu},
  \citenamefont {Liu}, \citenamefont {Rhodes}, \citenamefont {Watanabe},
  \citenamefont {Taniguchi}, \citenamefont {Hone}, \citenamefont {Elser},
  \citenamefont {Mak},\ and\ \citenamefont {Shan}}]{Xu20nat}%
  \BibitemOpen
  \bibfield  {author} {\bibinfo {author} {\bibfnamefont {Y.}~\bibnamefont
  {Xu}}, \bibinfo {author} {\bibfnamefont {S.}~\bibnamefont {Liu}}, \bibinfo
  {author} {\bibfnamefont {D.~A.}\ \bibnamefont {Rhodes}}, \bibinfo {author}
  {\bibfnamefont {K.}~\bibnamefont {Watanabe}}, \bibinfo {author}
  {\bibfnamefont {T.}~\bibnamefont {Taniguchi}}, \bibinfo {author}
  {\bibfnamefont {J.}~\bibnamefont {Hone}}, \bibinfo {author} {\bibfnamefont
  {V.}~\bibnamefont {Elser}}, \bibinfo {author} {\bibfnamefont {K.~F.}\
  \bibnamefont {Mak}},\ and\ \bibinfo {author} {\bibfnamefont {J.}~\bibnamefont
  {Shan}},\ }\bibfield  {title} {\bibinfo {title} {Correlated insulating states
  at fractional fillings of moiré superlattices},\ }\href
  {https://doi.org/10.1038/s41586-020-2868-6} {\bibfield  {journal} {\bibinfo
  {journal} {Nature}\ }\textbf {\bibinfo {volume} {587}},\ \bibinfo {pages}
  {214} (\bibinfo {year} {2020})}\BibitemShut {NoStop}%
\bibitem [{\citenamefont {Jin}\ \emph {et~al.}(2018)\citenamefont {Jin},
  \citenamefont {Ma}, \citenamefont {Karni}, \citenamefont {Regan},
  \citenamefont {Wang},\ and\ \citenamefont {Heinz}}]{Jin18natnano}%
  \BibitemOpen
  \bibfield  {author} {\bibinfo {author} {\bibfnamefont {C.}~\bibnamefont
  {Jin}}, \bibinfo {author} {\bibfnamefont {E.~Y.}\ \bibnamefont {Ma}},
  \bibinfo {author} {\bibfnamefont {O.}~\bibnamefont {Karni}}, \bibinfo
  {author} {\bibfnamefont {E.~C.}\ \bibnamefont {Regan}}, \bibinfo {author}
  {\bibfnamefont {F.}~\bibnamefont {Wang}},\ and\ \bibinfo {author}
  {\bibfnamefont {T.~F.}\ \bibnamefont {Heinz}},\ }\bibfield  {title} {\bibinfo
  {title} {{Ultrafast dynamics in van der {Waals} heterostructures}},\ }\href
  {https://doi.org/10.1038/s41565-018-0298-5} {\bibfield  {journal} {\bibinfo
  {journal} {Nature Nanotechnology}\ }\textbf {\bibinfo {volume} {13}},\
  \bibinfo {pages} {994} (\bibinfo {year} {2018})}\BibitemShut {NoStop}%
\bibitem [{\citenamefont {Zhang}\ \emph {et~al.}(2015)\citenamefont {Zhang},
  \citenamefont {You}, \citenamefont {Zhao},\ and\ \citenamefont
  {Heinz}}]{Zhang15prl}%
  \BibitemOpen
  \bibfield  {author} {\bibinfo {author} {\bibfnamefont {X.-X.}\ \bibnamefont
  {Zhang}}, \bibinfo {author} {\bibfnamefont {Y.}~\bibnamefont {You}}, \bibinfo
  {author} {\bibfnamefont {S.~Y.~F.}\ \bibnamefont {Zhao}},\ and\ \bibinfo
  {author} {\bibfnamefont {T.~F.}\ \bibnamefont {Heinz}},\ }\bibfield  {title}
  {\bibinfo {title} {Experimental evidence for dark excitons in monolayer
  {${\mathrm{WSe}}_{2}$}},\ }\href
  {https://doi.org/10.1103/PhysRevLett.115.257403} {\bibfield  {journal}
  {\bibinfo  {journal} {Physical Review Letters}\ }\textbf {\bibinfo {volume}
  {115}},\ \bibinfo {pages} {257403} (\bibinfo {year} {2015})}\BibitemShut
  {NoStop}%
\bibitem [{\citenamefont {Raja}\ \emph {et~al.}(2018)\citenamefont {Raja},
  \citenamefont {Selig}, \citenamefont {Berghäuser}, \citenamefont {Yu},
  \citenamefont {Hill}, \citenamefont {Rigosi}, \citenamefont {Brus},
  \citenamefont {Knorr}, \citenamefont {Heinz}, \citenamefont {Malic},\ and\
  \citenamefont {Chernikov}}]{Raja18nanolett}%
  \BibitemOpen
  \bibfield  {author} {\bibinfo {author} {\bibfnamefont {A.}~\bibnamefont
  {Raja}}, \bibinfo {author} {\bibfnamefont {M.}~\bibnamefont {Selig}},
  \bibinfo {author} {\bibfnamefont {G.}~\bibnamefont {Berghäuser}}, \bibinfo
  {author} {\bibfnamefont {J.}~\bibnamefont {Yu}}, \bibinfo {author}
  {\bibfnamefont {H.~M.}\ \bibnamefont {Hill}}, \bibinfo {author}
  {\bibfnamefont {A.~F.}\ \bibnamefont {Rigosi}}, \bibinfo {author}
  {\bibfnamefont {L.~E.}\ \bibnamefont {Brus}}, \bibinfo {author}
  {\bibfnamefont {A.}~\bibnamefont {Knorr}}, \bibinfo {author} {\bibfnamefont
  {T.~F.}\ \bibnamefont {Heinz}}, \bibinfo {author} {\bibfnamefont
  {E.}~\bibnamefont {Malic}},\ and\ \bibinfo {author} {\bibfnamefont
  {A.}~\bibnamefont {Chernikov}},\ }\bibfield  {title} {\bibinfo {title}
  {Enhancement of exciton–phonon scattering from monolayer to bilayer
  {WS$_2$}},\ }\href {https://doi.org/10.1021/acs.nanolett.8b01793} {\bibfield
  {journal} {\bibinfo  {journal} {Nano Letters}\ }\textbf {\bibinfo {volume}
  {18}},\ \bibinfo {pages} {6135} (\bibinfo {year} {2018})}\BibitemShut
  {NoStop}%
\bibitem [{\citenamefont {Jiang}\ \emph {et~al.}(2021)\citenamefont {Jiang},
  \citenamefont {Zheng}, \citenamefont {Lan}, \citenamefont {Saidi},
  \citenamefont {Ren},\ and\ \citenamefont {Zhao}}]{Jiang21sciadv}%
  \BibitemOpen
  \bibfield  {author} {\bibinfo {author} {\bibfnamefont {X.}~\bibnamefont
  {Jiang}}, \bibinfo {author} {\bibfnamefont {Q.}~\bibnamefont {Zheng}},
  \bibinfo {author} {\bibfnamefont {Z.}~\bibnamefont {Lan}}, \bibinfo {author}
  {\bibfnamefont {W.~A.}\ \bibnamefont {Saidi}}, \bibinfo {author}
  {\bibfnamefont {X.}~\bibnamefont {Ren}},\ and\ \bibinfo {author}
  {\bibfnamefont {J.}~\bibnamefont {Zhao}},\ }\bibfield  {title} {\bibinfo
  {title} {Real-time \textit{GW}-{BSE} investigations on spin-valley exciton
  dynamics in monolayer transition metal dichalcogenide},\ }\href
  {https://doi.org/doi:10.1126/sciadv.abf3759} {\bibfield  {journal} {\bibinfo
  {journal} {Science Advances}\ }\textbf {\bibinfo {volume} {7}},\ \bibinfo
  {pages} {eabf3759} (\bibinfo {year} {2021})}\BibitemShut {NoStop}%
\bibitem [{\citenamefont {Selig}\ \emph {et~al.}(2018)\citenamefont {Selig},
  \citenamefont {Berghäuser}, \citenamefont {Richter}, \citenamefont
  {Bratschitsch}, \citenamefont {Knorr},\ and\ \citenamefont
  {Malic}}]{Selig182Dmat}%
  \BibitemOpen
  \bibfield  {author} {\bibinfo {author} {\bibfnamefont {M.}~\bibnamefont
  {Selig}}, \bibinfo {author} {\bibfnamefont {G.}~\bibnamefont {Berghäuser}},
  \bibinfo {author} {\bibfnamefont {M.}~\bibnamefont {Richter}}, \bibinfo
  {author} {\bibfnamefont {R.}~\bibnamefont {Bratschitsch}}, \bibinfo {author}
  {\bibfnamefont {A.}~\bibnamefont {Knorr}},\ and\ \bibinfo {author}
  {\bibfnamefont {E.}~\bibnamefont {Malic}},\ }\bibfield  {title} {\bibinfo
  {title} {Dark and bright exciton formation, thermalization, and
  photoluminescence in monolayer transition metal dichalcogenides},\ }\href
  {https://doi.org/10.1088/2053-1583/aabea3} {\bibfield  {journal} {\bibinfo
  {journal} {2D Materials}\ }\textbf {\bibinfo {volume} {5}},\ \bibinfo {pages}
  {035017} (\bibinfo {year} {2018})}\BibitemShut {NoStop}%
\bibitem [{\citenamefont {Merkl}\ \emph {et~al.}(2019)\citenamefont {Merkl},
  \citenamefont {Mooshammer}, \citenamefont {Steinleitner}, \citenamefont
  {Girnghuber}, \citenamefont {Lin}, \citenamefont {Nagler}, \citenamefont
  {Holler}, \citenamefont {Schüller}, \citenamefont {Lupton}, \citenamefont
  {Korn}, \citenamefont {Ovesen}, \citenamefont {Brem}, \citenamefont {Malic},\
  and\ \citenamefont {Huber}}]{Merkl19natmat}%
  \BibitemOpen
  \bibfield  {author} {\bibinfo {author} {\bibfnamefont {P.}~\bibnamefont
  {Merkl}}, \bibinfo {author} {\bibfnamefont {F.}~\bibnamefont {Mooshammer}},
  \bibinfo {author} {\bibfnamefont {P.}~\bibnamefont {Steinleitner}}, \bibinfo
  {author} {\bibfnamefont {A.}~\bibnamefont {Girnghuber}}, \bibinfo {author}
  {\bibfnamefont {K.~Q.}\ \bibnamefont {Lin}}, \bibinfo {author} {\bibfnamefont
  {P.}~\bibnamefont {Nagler}}, \bibinfo {author} {\bibfnamefont
  {J.}~\bibnamefont {Holler}}, \bibinfo {author} {\bibfnamefont
  {C.}~\bibnamefont {Schüller}}, \bibinfo {author} {\bibfnamefont {J.~M.}\
  \bibnamefont {Lupton}}, \bibinfo {author} {\bibfnamefont {T.}~\bibnamefont
  {Korn}}, \bibinfo {author} {\bibfnamefont {S.}~\bibnamefont {Ovesen}},
  \bibinfo {author} {\bibfnamefont {S.}~\bibnamefont {Brem}}, \bibinfo {author}
  {\bibfnamefont {E.}~\bibnamefont {Malic}},\ and\ \bibinfo {author}
  {\bibfnamefont {R.}~\bibnamefont {Huber}},\ }\bibfield  {title} {\bibinfo
  {title} {{Ultrafast transition between exciton phases in van der Waals
  heterostructures}},\ }\href {https://doi.org/10.1038/s41563-019-0337-0}
  {\bibfield  {journal} {\bibinfo  {journal} {Nature Materials}\ }\textbf
  {\bibinfo {volume} {18}},\ \bibinfo {pages} {691} (\bibinfo {year}
  {2019})}\BibitemShut {NoStop}%
\bibitem [{\citenamefont {Madéo}\ \emph {et~al.}(2020)\citenamefont {Madéo},
  \citenamefont {Man}, \citenamefont {Sahoo}, \citenamefont {Campbell},
  \citenamefont {Pareek}, \citenamefont {Wong}, \citenamefont {Al-Mahboob},
  \citenamefont {Chan}, \citenamefont {Karmakar}, \citenamefont {Mariserla},
  \citenamefont {Li}, \citenamefont {Heinz}, \citenamefont {Cao},\ and\
  \citenamefont {Dani}}]{Madeo20sci}%
  \BibitemOpen
  \bibfield  {author} {\bibinfo {author} {\bibfnamefont {J.}~\bibnamefont
  {Madéo}}, \bibinfo {author} {\bibfnamefont {M.~K.~L.}\ \bibnamefont {Man}},
  \bibinfo {author} {\bibfnamefont {C.}~\bibnamefont {Sahoo}}, \bibinfo
  {author} {\bibfnamefont {M.}~\bibnamefont {Campbell}}, \bibinfo {author}
  {\bibfnamefont {V.}~\bibnamefont {Pareek}}, \bibinfo {author} {\bibfnamefont
  {E.~L.}\ \bibnamefont {Wong}}, \bibinfo {author} {\bibfnamefont
  {A.}~\bibnamefont {Al-Mahboob}}, \bibinfo {author} {\bibfnamefont {N.~S.}\
  \bibnamefont {Chan}}, \bibinfo {author} {\bibfnamefont {A.}~\bibnamefont
  {Karmakar}}, \bibinfo {author} {\bibfnamefont {B.~M.~K.}\ \bibnamefont
  {Mariserla}}, \bibinfo {author} {\bibfnamefont {X.}~\bibnamefont {Li}},
  \bibinfo {author} {\bibfnamefont {T.~F.}\ \bibnamefont {Heinz}}, \bibinfo
  {author} {\bibfnamefont {T.}~\bibnamefont {Cao}},\ and\ \bibinfo {author}
  {\bibfnamefont {K.~M.}\ \bibnamefont {Dani}},\ }\bibfield  {title} {\bibinfo
  {title} {{Directly visualizing the momentum-forbidden dark excitons and their
  dynamics in atomically thin semiconductors}},\ }\href
  {https://doi.org/10.1126/science.aba1029} {\bibfield  {journal} {\bibinfo
  {journal} {Science}\ }\textbf {\bibinfo {volume} {370}},\ \bibinfo {pages}
  {1199} (\bibinfo {year} {2020})}\BibitemShut {NoStop}%
\bibitem [{\citenamefont {Wallauer}\ \emph {et~al.}(2021)\citenamefont
  {Wallauer}, \citenamefont {Perea-Causin}, \citenamefont {Münster},
  \citenamefont {Zajusch}, \citenamefont {Brem}, \citenamefont {Güdde},
  \citenamefont {Tanimura}, \citenamefont {Lin}, \citenamefont {Huber},
  \citenamefont {Malic},\ and\ \citenamefont {Höfer}}]{Wallauer21nanolett}%
  \BibitemOpen
  \bibfield  {author} {\bibinfo {author} {\bibfnamefont {R.}~\bibnamefont
  {Wallauer}}, \bibinfo {author} {\bibfnamefont {R.}~\bibnamefont
  {Perea-Causin}}, \bibinfo {author} {\bibfnamefont {L.}~\bibnamefont
  {Münster}}, \bibinfo {author} {\bibfnamefont {S.}~\bibnamefont {Zajusch}},
  \bibinfo {author} {\bibfnamefont {S.}~\bibnamefont {Brem}}, \bibinfo {author}
  {\bibfnamefont {J.}~\bibnamefont {Güdde}}, \bibinfo {author} {\bibfnamefont
  {K.}~\bibnamefont {Tanimura}}, \bibinfo {author} {\bibfnamefont {K.-Q.}\
  \bibnamefont {Lin}}, \bibinfo {author} {\bibfnamefont {R.}~\bibnamefont
  {Huber}}, \bibinfo {author} {\bibfnamefont {E.}~\bibnamefont {Malic}},\ and\
  \bibinfo {author} {\bibfnamefont {U.}~\bibnamefont {Höfer}},\ }\bibfield
  {title} {\bibinfo {title} {Momentum-resolved observation of exciton formation
  dynamics in monolayer {WS}$_2$},\ }\href@noop {} {\bibfield  {journal}
  {\bibinfo  {journal} {Nano Letters}\ }\textbf {\bibinfo {volume} {21}},\
  \bibinfo {pages} {5867} (\bibinfo {year} {2021})}\BibitemShut {NoStop}%
\bibitem [{\citenamefont {Bange}\ \emph {et~al.}(2023)\citenamefont {Bange},
  \citenamefont {Werner}, \citenamefont {Schmitt}, \citenamefont {Bennecke},
  \citenamefont {Meneghini}, \citenamefont {AlMutairi}, \citenamefont
  {Merboldt}, \citenamefont {Watanabe}, \citenamefont {Taniguchi},
  \citenamefont {Steil}, \citenamefont {Steil}, \citenamefont {Weitz},
  \citenamefont {Hofmann}, \citenamefont {Jansen}, \citenamefont {Brem},
  \citenamefont {Malic}, \citenamefont {Reutzel},\ and\ \citenamefont
  {Mathias}}]{Bange232DMaterials}%
  \BibitemOpen
  \bibfield  {author} {\bibinfo {author} {\bibfnamefont {J.~P.}\ \bibnamefont
  {Bange}}, \bibinfo {author} {\bibfnamefont {P.}~\bibnamefont {Werner}},
  \bibinfo {author} {\bibfnamefont {D.}~\bibnamefont {Schmitt}}, \bibinfo
  {author} {\bibfnamefont {W.}~\bibnamefont {Bennecke}}, \bibinfo {author}
  {\bibfnamefont {G.}~\bibnamefont {Meneghini}}, \bibinfo {author}
  {\bibfnamefont {A.}~\bibnamefont {AlMutairi}}, \bibinfo {author}
  {\bibfnamefont {M.}~\bibnamefont {Merboldt}}, \bibinfo {author}
  {\bibfnamefont {K.}~\bibnamefont {Watanabe}}, \bibinfo {author}
  {\bibfnamefont {T.}~\bibnamefont {Taniguchi}}, \bibinfo {author}
  {\bibfnamefont {S.}~\bibnamefont {Steil}}, \bibinfo {author} {\bibfnamefont
  {D.}~\bibnamefont {Steil}}, \bibinfo {author} {\bibfnamefont {R.~T.}\
  \bibnamefont {Weitz}}, \bibinfo {author} {\bibfnamefont {S.}~\bibnamefont
  {Hofmann}}, \bibinfo {author} {\bibfnamefont {G.~S.~M.}\ \bibnamefont
  {Jansen}}, \bibinfo {author} {\bibfnamefont {S.}~\bibnamefont {Brem}},
  \bibinfo {author} {\bibfnamefont {E.}~\bibnamefont {Malic}}, \bibinfo
  {author} {\bibfnamefont {M.}~\bibnamefont {Reutzel}},\ and\ \bibinfo {author}
  {\bibfnamefont {S.}~\bibnamefont {Mathias}},\ }\bibfield  {title} {\bibinfo
  {title} {Ultrafast dynamics of bright and dark excitons in monolayer
  {WSe$_2$} and heterobilayer {WSe$_2$/MoS$_2$}},\ }\href
  {https://doi.org/10.1088/2053-1583/ace067} {\bibfield  {journal} {\bibinfo
  {journal} {2D Materials}\ }\textbf {\bibinfo {volume} {10}},\ \bibinfo
  {pages} {035039} (\bibinfo {year} {2023})}\BibitemShut {NoStop}%
\bibitem [{\citenamefont {Caruso}(2021)}]{Caruso21jpcl}%
  \BibitemOpen
  \bibfield  {author} {\bibinfo {author} {\bibfnamefont {F.}~\bibnamefont
  {Caruso}},\ }\bibfield  {title} {\bibinfo {title} {Nonequilibrium lattice
  dynamics in monolayer {MoS$_2$}},\ }\href
  {https://doi.org/10.1021/acs.jpclett.0c03616} {\bibfield  {journal} {\bibinfo
   {journal} {The Journal of Physical Chemistry Letters}\ }\textbf {\bibinfo
  {volume} {12}},\ \bibinfo {pages} {1734} (\bibinfo {year}
  {2021})}\BibitemShut {NoStop}%
\bibitem [{\citenamefont {Ovesen}\ \emph {et~al.}(2019)\citenamefont {Ovesen},
  \citenamefont {Brem}, \citenamefont {Linderälv}, \citenamefont {Kuisma},
  \citenamefont {Korn}, \citenamefont {Erhart}, \citenamefont {Selig},\ and\
  \citenamefont {Malic}}]{Ovesen19comphys}%
  \BibitemOpen
  \bibfield  {author} {\bibinfo {author} {\bibfnamefont {S.}~\bibnamefont
  {Ovesen}}, \bibinfo {author} {\bibfnamefont {S.}~\bibnamefont {Brem}},
  \bibinfo {author} {\bibfnamefont {C.}~\bibnamefont {Linderälv}}, \bibinfo
  {author} {\bibfnamefont {M.}~\bibnamefont {Kuisma}}, \bibinfo {author}
  {\bibfnamefont {T.}~\bibnamefont {Korn}}, \bibinfo {author} {\bibfnamefont
  {P.}~\bibnamefont {Erhart}}, \bibinfo {author} {\bibfnamefont
  {M.}~\bibnamefont {Selig}},\ and\ \bibinfo {author} {\bibfnamefont
  {E.}~\bibnamefont {Malic}},\ }\bibfield  {title} {\bibinfo {title}
  {{Interlayer exciton dynamics in van der Waals heterostructures}},\ }\href
  {https://doi.org/10.1038/s42005-019-0122-z} {\bibfield  {journal} {\bibinfo
  {journal} {Communications Physics}\ }\textbf {\bibinfo {volume} {2}},\
  \bibinfo {pages} {23} (\bibinfo {year} {2019})}\BibitemShut {NoStop}%
\bibitem [{\citenamefont {Meneghini}\ \emph {et~al.}(2023)\citenamefont
  {Meneghini}, \citenamefont {Reutzel}, \citenamefont {Mathias}, \citenamefont
  {Brem},\ and\ \citenamefont {Malic}}]{Meneghini23acspho}%
  \BibitemOpen
  \bibfield  {author} {\bibinfo {author} {\bibfnamefont {G.}~\bibnamefont
  {Meneghini}}, \bibinfo {author} {\bibfnamefont {M.}~\bibnamefont {Reutzel}},
  \bibinfo {author} {\bibfnamefont {S.}~\bibnamefont {Mathias}}, \bibinfo
  {author} {\bibfnamefont {S.}~\bibnamefont {Brem}},\ and\ \bibinfo {author}
  {\bibfnamefont {E.}~\bibnamefont {Malic}},\ }\bibfield  {title} {\bibinfo
  {title} {Hybrid exciton signatures in arpes spectra of van der waals
  materials},\ }\href {https://doi.org/10.1021/acsphotonics.3c00599} {\bibfield
   {journal} {\bibinfo  {journal} {ACS Photonics}\ }\textbf {\bibinfo {volume}
  {10}},\ \bibinfo {pages} {3570} (\bibinfo {year} {2023})}\BibitemShut
  {NoStop}%
\bibitem [{\citenamefont {Policht}\ \emph {et~al.}(2023)\citenamefont
  {Policht}, \citenamefont {Mittenzwey}, \citenamefont {Dogadov}, \citenamefont
  {Katzer}, \citenamefont {Villa}, \citenamefont {Li}, \citenamefont {Kaiser},
  \citenamefont {Ross}, \citenamefont {Scotognella}, \citenamefont {Zhu},
  \citenamefont {Knorr}, \citenamefont {Selig}, \citenamefont {Cerullo},\ and\
  \citenamefont {Dal~Conte}}]{Policht23natcom}%
  \BibitemOpen
  \bibfield  {author} {\bibinfo {author} {\bibfnamefont {V.~R.}\ \bibnamefont
  {Policht}}, \bibinfo {author} {\bibfnamefont {H.}~\bibnamefont {Mittenzwey}},
  \bibinfo {author} {\bibfnamefont {O.}~\bibnamefont {Dogadov}}, \bibinfo
  {author} {\bibfnamefont {M.}~\bibnamefont {Katzer}}, \bibinfo {author}
  {\bibfnamefont {A.}~\bibnamefont {Villa}}, \bibinfo {author} {\bibfnamefont
  {Q.}~\bibnamefont {Li}}, \bibinfo {author} {\bibfnamefont {B.}~\bibnamefont
  {Kaiser}}, \bibinfo {author} {\bibfnamefont {A.~M.}\ \bibnamefont {Ross}},
  \bibinfo {author} {\bibfnamefont {F.}~\bibnamefont {Scotognella}}, \bibinfo
  {author} {\bibfnamefont {X.}~\bibnamefont {Zhu}}, \bibinfo {author}
  {\bibfnamefont {A.}~\bibnamefont {Knorr}}, \bibinfo {author} {\bibfnamefont
  {M.}~\bibnamefont {Selig}}, \bibinfo {author} {\bibfnamefont
  {G.}~\bibnamefont {Cerullo}},\ and\ \bibinfo {author} {\bibfnamefont
  {S.}~\bibnamefont {Dal~Conte}},\ }\bibfield  {title} {\bibinfo {title}
  {Time-domain observation of interlayer exciton formation and thermalization
  in a {MoSe$_2$/WSe$_2$} heterostructure},\ }\href
  {https://doi.org/10.1038/s41467-023-42915-x} {\bibfield  {journal} {\bibinfo
  {journal} {Nature Communications}\ }\textbf {\bibinfo {volume} {14}},\
  \bibinfo {pages} {7273} (\bibinfo {year} {2023})}\BibitemShut {NoStop}%
\bibitem [{\citenamefont {Rosati}\ \emph {et~al.}(2020)\citenamefont {Rosati},
  \citenamefont {Wagner}, \citenamefont {Brem}, \citenamefont {Perea-Causín},
  \citenamefont {Wietek}, \citenamefont {Zipfel}, \citenamefont {Ziegler},
  \citenamefont {Selig}, \citenamefont {Taniguchi}, \citenamefont {Watanabe},
  \citenamefont {Knorr}, \citenamefont {Chernikov},\ and\ \citenamefont
  {Malic}}]{Rosati20acsphotonics}%
  \BibitemOpen
  \bibfield  {author} {\bibinfo {author} {\bibfnamefont {R.}~\bibnamefont
  {Rosati}}, \bibinfo {author} {\bibfnamefont {K.}~\bibnamefont {Wagner}},
  \bibinfo {author} {\bibfnamefont {S.}~\bibnamefont {Brem}}, \bibinfo {author}
  {\bibfnamefont {R.}~\bibnamefont {Perea-Causín}}, \bibinfo {author}
  {\bibfnamefont {E.}~\bibnamefont {Wietek}}, \bibinfo {author} {\bibfnamefont
  {J.}~\bibnamefont {Zipfel}}, \bibinfo {author} {\bibfnamefont {J.~D.}\
  \bibnamefont {Ziegler}}, \bibinfo {author} {\bibfnamefont {M.}~\bibnamefont
  {Selig}}, \bibinfo {author} {\bibfnamefont {T.}~\bibnamefont {Taniguchi}},
  \bibinfo {author} {\bibfnamefont {K.}~\bibnamefont {Watanabe}}, \bibinfo
  {author} {\bibfnamefont {A.}~\bibnamefont {Knorr}}, \bibinfo {author}
  {\bibfnamefont {A.}~\bibnamefont {Chernikov}},\ and\ \bibinfo {author}
  {\bibfnamefont {E.}~\bibnamefont {Malic}},\ }\bibfield  {title} {\bibinfo
  {title} {Temporal evolution of low-temperature phonon sidebands in transition
  metal dichalcogenides},\ }\href
  {https://doi.org/10.1021/acsphotonics.0c00866} {\bibfield  {journal}
  {\bibinfo  {journal} {ACS Photonics}\ }\textbf {\bibinfo {volume} {7}},\
  \bibinfo {pages} {2756} (\bibinfo {year} {2020})}\BibitemShut {NoStop}%
\bibitem [{\citenamefont {Poellmann}\ \emph {et~al.}(2015)\citenamefont
  {Poellmann}, \citenamefont {Steinleitner}, \citenamefont {Leierseder},
  \citenamefont {Nagler}, \citenamefont {Plechinger}, \citenamefont {Porer},
  \citenamefont {Bratschitsch}, \citenamefont {Schüller}, \citenamefont
  {Korn},\ and\ \citenamefont {Huber}}]{Poellmann15natmat}%
  \BibitemOpen
  \bibfield  {author} {\bibinfo {author} {\bibfnamefont {C.}~\bibnamefont
  {Poellmann}}, \bibinfo {author} {\bibfnamefont {P.}~\bibnamefont
  {Steinleitner}}, \bibinfo {author} {\bibfnamefont {U.}~\bibnamefont
  {Leierseder}}, \bibinfo {author} {\bibfnamefont {P.}~\bibnamefont {Nagler}},
  \bibinfo {author} {\bibfnamefont {G.}~\bibnamefont {Plechinger}}, \bibinfo
  {author} {\bibfnamefont {M.}~\bibnamefont {Porer}}, \bibinfo {author}
  {\bibfnamefont {R.}~\bibnamefont {Bratschitsch}}, \bibinfo {author}
  {\bibfnamefont {C.}~\bibnamefont {Schüller}}, \bibinfo {author}
  {\bibfnamefont {T.}~\bibnamefont {Korn}},\ and\ \bibinfo {author}
  {\bibfnamefont {R.}~\bibnamefont {Huber}},\ }\bibfield  {title} {\bibinfo
  {title} {Resonant internal quantum transitions and femtosecond radiative
  decay of excitons in monolayer {WSe$_2$}},\ }\href
  {https://doi.org/10.1038/nmat4356} {\bibfield  {journal} {\bibinfo  {journal}
  {Nature Materials}\ }\textbf {\bibinfo {volume} {14}},\ \bibinfo {pages}
  {889} (\bibinfo {year} {2015})}\BibitemShut {NoStop}%
\bibitem [{\citenamefont {Reutzel}\ \emph {et~al.}(2024)\citenamefont
  {Reutzel}, \citenamefont {Jansen},\ and\ \citenamefont
  {Mathias}}]{Reutzel24AdvPhysX}%
  \BibitemOpen
  \bibfield  {author} {\bibinfo {author} {\bibfnamefont {M.}~\bibnamefont
  {Reutzel}}, \bibinfo {author} {\bibfnamefont {G.~S.~M.}\ \bibnamefont
  {Jansen}},\ and\ \bibinfo {author} {\bibfnamefont {S.}~\bibnamefont
  {Mathias}},\ }\bibfield  {title} {\bibinfo {title} {Probing excitons with
  time-resolved momentum microscopy},\ }\href
  {https://doi.org/10.1080/23746149.2024.2378722} {\bibfield  {journal}
  {\bibinfo  {journal} {Advances in Physics: X}\ }\textbf {\bibinfo {volume}
  {9}},\ \bibinfo {pages} {2378722} (\bibinfo {year} {2024})}\BibitemShut
  {NoStop}%
\bibitem [{\citenamefont {Dong}\ \emph {et~al.}(2021)\citenamefont {Dong},
  \citenamefont {Puppin}, \citenamefont {Pincelli}, \citenamefont {Beaulieu},
  \citenamefont {Christiansen}, \citenamefont {Hübener}, \citenamefont
  {Nicholson}, \citenamefont {Xian}, \citenamefont {Dendzik}, \citenamefont
  {Deng}, \citenamefont {Windsor}, \citenamefont {Selig}, \citenamefont
  {Malic}, \citenamefont {Rubio}, \citenamefont {Knorr}, \citenamefont {Wolf},
  \citenamefont {Rettig},\ and\ \citenamefont
  {Ernstorfer}}]{Dong20naturalsciences}%
  \BibitemOpen
  \bibfield  {author} {\bibinfo {author} {\bibfnamefont {S.}~\bibnamefont
  {Dong}}, \bibinfo {author} {\bibfnamefont {M.}~\bibnamefont {Puppin}},
  \bibinfo {author} {\bibfnamefont {T.}~\bibnamefont {Pincelli}}, \bibinfo
  {author} {\bibfnamefont {S.}~\bibnamefont {Beaulieu}}, \bibinfo {author}
  {\bibfnamefont {D.}~\bibnamefont {Christiansen}}, \bibinfo {author}
  {\bibfnamefont {H.}~\bibnamefont {Hübener}}, \bibinfo {author}
  {\bibfnamefont {C.~W.}\ \bibnamefont {Nicholson}}, \bibinfo {author}
  {\bibfnamefont {R.~P.}\ \bibnamefont {Xian}}, \bibinfo {author}
  {\bibfnamefont {M.}~\bibnamefont {Dendzik}}, \bibinfo {author} {\bibfnamefont
  {Y.}~\bibnamefont {Deng}}, \bibinfo {author} {\bibfnamefont {Y.~W.}\
  \bibnamefont {Windsor}}, \bibinfo {author} {\bibfnamefont {M.}~\bibnamefont
  {Selig}}, \bibinfo {author} {\bibfnamefont {E.}~\bibnamefont {Malic}},
  \bibinfo {author} {\bibfnamefont {A.}~\bibnamefont {Rubio}}, \bibinfo
  {author} {\bibfnamefont {A.}~\bibnamefont {Knorr}}, \bibinfo {author}
  {\bibfnamefont {M.}~\bibnamefont {Wolf}}, \bibinfo {author} {\bibfnamefont
  {L.}~\bibnamefont {Rettig}},\ and\ \bibinfo {author} {\bibfnamefont
  {R.}~\bibnamefont {Ernstorfer}},\ }\bibfield  {title} {\bibinfo {title}
  {{Direct measurement of key exciton properties: Energy, dynamics, and spatial
  distribution of the wave function}},\ }\href
  {https://doi.org/https://doi.org/10.1002/ntls.10010} {\bibfield  {journal}
  {\bibinfo  {journal} {Natural Sciences}\ }\textbf {\bibinfo {volume} {1}},\
  \bibinfo {pages} {e10010} (\bibinfo {year} {2021})}\BibitemShut {NoStop}%
\bibitem [{\citenamefont {Man}\ \emph {et~al.}(2021)\citenamefont {Man},
  \citenamefont {Madéo}, \citenamefont {Sahoo}, \citenamefont {Xie},
  \citenamefont {Campbell}, \citenamefont {Pareek}, \citenamefont {Karmakar},
  \citenamefont {Wong}, \citenamefont {Al-Mahboob}, \citenamefont {Chan},
  \citenamefont {Bacon}, \citenamefont {Zhu}, \citenamefont {Abdelrasoul},
  \citenamefont {Li}, \citenamefont {Heinz}, \citenamefont {da~Jornada},
  \citenamefont {Cao},\ and\ \citenamefont {Dani}}]{Man21sciadv}%
  \BibitemOpen
  \bibfield  {author} {\bibinfo {author} {\bibfnamefont {M.~K.~L.}\
  \bibnamefont {Man}}, \bibinfo {author} {\bibfnamefont {J.}~\bibnamefont
  {Madéo}}, \bibinfo {author} {\bibfnamefont {C.}~\bibnamefont {Sahoo}},
  \bibinfo {author} {\bibfnamefont {K.}~\bibnamefont {Xie}}, \bibinfo {author}
  {\bibfnamefont {M.}~\bibnamefont {Campbell}}, \bibinfo {author}
  {\bibfnamefont {V.}~\bibnamefont {Pareek}}, \bibinfo {author} {\bibfnamefont
  {A.}~\bibnamefont {Karmakar}}, \bibinfo {author} {\bibfnamefont {E.~L.}\
  \bibnamefont {Wong}}, \bibinfo {author} {\bibfnamefont {A.}~\bibnamefont
  {Al-Mahboob}}, \bibinfo {author} {\bibfnamefont {N.~S.}\ \bibnamefont
  {Chan}}, \bibinfo {author} {\bibfnamefont {D.~R.}\ \bibnamefont {Bacon}},
  \bibinfo {author} {\bibfnamefont {X.}~\bibnamefont {Zhu}}, \bibinfo {author}
  {\bibfnamefont {M.~M.~M.}\ \bibnamefont {Abdelrasoul}}, \bibinfo {author}
  {\bibfnamefont {X.}~\bibnamefont {Li}}, \bibinfo {author} {\bibfnamefont
  {T.~F.}\ \bibnamefont {Heinz}}, \bibinfo {author} {\bibfnamefont {F.~H.}\
  \bibnamefont {da~Jornada}}, \bibinfo {author} {\bibfnamefont
  {T.}~\bibnamefont {Cao}},\ and\ \bibinfo {author} {\bibfnamefont {K.~M.}\
  \bibnamefont {Dani}},\ }\bibfield  {title} {\bibinfo {title} {{Experimental
  measurement of the intrinsic excitonic wave function}},\ }\href
  {https://doi.org/10.1126/sciadv.abg0192} {\bibfield  {journal} {\bibinfo
  {journal} {Science Advances}\ }\textbf {\bibinfo {volume} {7}},\ \bibinfo
  {pages} {eabg0192} (\bibinfo {year} {2021})}\BibitemShut {NoStop}%
\bibitem [{\citenamefont {Schmitt}\ \emph {et~al.}(2022)\citenamefont
  {Schmitt}, \citenamefont {Bange}, \citenamefont {Bennecke}, \citenamefont
  {AlMutairi}, \citenamefont {Meneghini}, \citenamefont {Watanabe},
  \citenamefont {Taniguchi}, \citenamefont {Steil}, \citenamefont {Luke},
  \citenamefont {Weitz}, \citenamefont {Steil}, \citenamefont {Jansen},
  \citenamefont {Brem}, \citenamefont {Malic}, \citenamefont {Hofmann},
  \citenamefont {Reutzel},\ and\ \citenamefont {Mathias}}]{Schmitt22nat}%
  \BibitemOpen
  \bibfield  {author} {\bibinfo {author} {\bibfnamefont {D.}~\bibnamefont
  {Schmitt}}, \bibinfo {author} {\bibfnamefont {J.~P.}\ \bibnamefont {Bange}},
  \bibinfo {author} {\bibfnamefont {W.}~\bibnamefont {Bennecke}}, \bibinfo
  {author} {\bibfnamefont {A.}~\bibnamefont {AlMutairi}}, \bibinfo {author}
  {\bibfnamefont {G.}~\bibnamefont {Meneghini}}, \bibinfo {author}
  {\bibfnamefont {K.}~\bibnamefont {Watanabe}}, \bibinfo {author}
  {\bibfnamefont {T.}~\bibnamefont {Taniguchi}}, \bibinfo {author}
  {\bibfnamefont {D.}~\bibnamefont {Steil}}, \bibinfo {author} {\bibfnamefont
  {D.~R.}\ \bibnamefont {Luke}}, \bibinfo {author} {\bibfnamefont {R.~T.}\
  \bibnamefont {Weitz}}, \bibinfo {author} {\bibfnamefont {S.}~\bibnamefont
  {Steil}}, \bibinfo {author} {\bibfnamefont {G.~S.~M.}\ \bibnamefont
  {Jansen}}, \bibinfo {author} {\bibfnamefont {S.}~\bibnamefont {Brem}},
  \bibinfo {author} {\bibfnamefont {E.}~\bibnamefont {Malic}}, \bibinfo
  {author} {\bibfnamefont {S.}~\bibnamefont {Hofmann}}, \bibinfo {author}
  {\bibfnamefont {M.}~\bibnamefont {Reutzel}},\ and\ \bibinfo {author}
  {\bibfnamefont {S.}~\bibnamefont {Mathias}},\ }\bibfield  {title} {\bibinfo
  {title} {Formation of moiré interlayer excitons in space and time},\ }\href
  {https://doi.org/10.1038/s41586-022-04977-7} {\bibfield  {journal} {\bibinfo
  {journal} {Nature}\ }\textbf {\bibinfo {volume} {608}},\ \bibinfo {pages}
  {499} (\bibinfo {year} {2022})}\BibitemShut {NoStop}%
\bibitem [{\citenamefont {Karni}\ \emph {et~al.}(2022)\citenamefont {Karni},
  \citenamefont {Barré}, \citenamefont {Pareek}, \citenamefont {Georgaras},
  \citenamefont {Man}, \citenamefont {Sahoo}, \citenamefont {Bacon},
  \citenamefont {Zhu}, \citenamefont {Ribeiro}, \citenamefont {O’Beirne},
  \citenamefont {Hu}, \citenamefont {Al-Mahboob}, \citenamefont {Abdelrasoul},
  \citenamefont {Chan}, \citenamefont {Karmakar}, \citenamefont {Winchester},
  \citenamefont {Kim}, \citenamefont {Watanabe}, \citenamefont {Taniguchi},
  \citenamefont {Barmak}, \citenamefont {Madéo}, \citenamefont {da~Jornada},
  \citenamefont {Heinz},\ and\ \citenamefont {Dani}}]{Karni22nat}%
  \BibitemOpen
  \bibfield  {author} {\bibinfo {author} {\bibfnamefont {O.}~\bibnamefont
  {Karni}}, \bibinfo {author} {\bibfnamefont {E.}~\bibnamefont {Barré}},
  \bibinfo {author} {\bibfnamefont {V.}~\bibnamefont {Pareek}}, \bibinfo
  {author} {\bibfnamefont {J.~D.}\ \bibnamefont {Georgaras}}, \bibinfo {author}
  {\bibfnamefont {M.~K.~L.}\ \bibnamefont {Man}}, \bibinfo {author}
  {\bibfnamefont {C.}~\bibnamefont {Sahoo}}, \bibinfo {author} {\bibfnamefont
  {D.~R.}\ \bibnamefont {Bacon}}, \bibinfo {author} {\bibfnamefont
  {X.}~\bibnamefont {Zhu}}, \bibinfo {author} {\bibfnamefont {H.~B.}\
  \bibnamefont {Ribeiro}}, \bibinfo {author} {\bibfnamefont {A.~L.}\
  \bibnamefont {O’Beirne}}, \bibinfo {author} {\bibfnamefont
  {J.}~\bibnamefont {Hu}}, \bibinfo {author} {\bibfnamefont {A.}~\bibnamefont
  {Al-Mahboob}}, \bibinfo {author} {\bibfnamefont {M.~M.~M.}\ \bibnamefont
  {Abdelrasoul}}, \bibinfo {author} {\bibfnamefont {N.~S.}\ \bibnamefont
  {Chan}}, \bibinfo {author} {\bibfnamefont {A.}~\bibnamefont {Karmakar}},
  \bibinfo {author} {\bibfnamefont {A.~J.}\ \bibnamefont {Winchester}},
  \bibinfo {author} {\bibfnamefont {B.}~\bibnamefont {Kim}}, \bibinfo {author}
  {\bibfnamefont {K.}~\bibnamefont {Watanabe}}, \bibinfo {author}
  {\bibfnamefont {T.}~\bibnamefont {Taniguchi}}, \bibinfo {author}
  {\bibfnamefont {K.}~\bibnamefont {Barmak}}, \bibinfo {author} {\bibfnamefont
  {J.}~\bibnamefont {Madéo}}, \bibinfo {author} {\bibfnamefont {F.~H.}\
  \bibnamefont {da~Jornada}}, \bibinfo {author} {\bibfnamefont {T.~F.}\
  \bibnamefont {Heinz}},\ and\ \bibinfo {author} {\bibfnamefont {K.~M.}\
  \bibnamefont {Dani}},\ }\bibfield  {title} {\bibinfo {title} {Structure of
  the moiré exciton captured by imaging its electron and hole},\ }\href
  {https://doi.org/10.1038/s41586-021-04360-y} {\bibfield  {journal} {\bibinfo
  {journal} {Nature}\ }\textbf {\bibinfo {volume} {603}},\ \bibinfo {pages}
  {247} (\bibinfo {year} {2022})}\BibitemShut {NoStop}%
\bibitem [{\citenamefont {Kunin}\ \emph {et~al.}(2023)\citenamefont {Kunin},
  \citenamefont {Chernov}, \citenamefont {Bakalis}, \citenamefont {Li},
  \citenamefont {Cheng}, \citenamefont {Withers}, \citenamefont {White},
  \citenamefont {Schönhense}, \citenamefont {Du}, \citenamefont {Kawakami},\
  and\ \citenamefont {Allison}}]{kunin23prl}%
  \BibitemOpen
  \bibfield  {author} {\bibinfo {author} {\bibfnamefont {A.}~\bibnamefont
  {Kunin}}, \bibinfo {author} {\bibfnamefont {S.}~\bibnamefont {Chernov}},
  \bibinfo {author} {\bibfnamefont {J.}~\bibnamefont {Bakalis}}, \bibinfo
  {author} {\bibfnamefont {Z.}~\bibnamefont {Li}}, \bibinfo {author}
  {\bibfnamefont {S.}~\bibnamefont {Cheng}}, \bibinfo {author} {\bibfnamefont
  {Z.~H.}\ \bibnamefont {Withers}}, \bibinfo {author} {\bibfnamefont {M.~G.}\
  \bibnamefont {White}}, \bibinfo {author} {\bibfnamefont {G.}~\bibnamefont
  {Schönhense}}, \bibinfo {author} {\bibfnamefont {X.}~\bibnamefont {Du}},
  \bibinfo {author} {\bibfnamefont {R.~K.}\ \bibnamefont {Kawakami}},\ and\
  \bibinfo {author} {\bibfnamefont {T.~K.}\ \bibnamefont {Allison}},\
  }\bibfield  {title} {\bibinfo {title} {Momentum-resolved exciton coupling and
  valley polarization dynamics in monolayer {${\mathrm{WS}}_{2}$}},\ }\href
  {https://doi.org/10.1103/PhysRevLett.130.046202} {\bibfield  {journal}
  {\bibinfo  {journal} {Physical Review Letters}\ }\textbf {\bibinfo {volume}
  {130}},\ \bibinfo {pages} {046202} (\bibinfo {year} {2023})}\BibitemShut
  {NoStop}%
\bibitem [{\citenamefont {Bange}\ \emph {et~al.}(2024)\citenamefont {Bange},
  \citenamefont {Schmitt}, \citenamefont {Bennecke}, \citenamefont {Meneghini},
  \citenamefont {AlMutairi}, \citenamefont {Watanabe}, \citenamefont
  {Taniguchi}, \citenamefont {Steil}, \citenamefont {Steil}, \citenamefont
  {Weitz}, \citenamefont {Jansen}, \citenamefont {Hofmann}, \citenamefont
  {Brem}, \citenamefont {Malic}, \citenamefont {Reutzel},\ and\ \citenamefont
  {Mathias}}]{Bange24SciAdv}%
  \BibitemOpen
  \bibfield  {author} {\bibinfo {author} {\bibfnamefont {J.~P.}\ \bibnamefont
  {Bange}}, \bibinfo {author} {\bibfnamefont {D.}~\bibnamefont {Schmitt}},
  \bibinfo {author} {\bibfnamefont {W.}~\bibnamefont {Bennecke}}, \bibinfo
  {author} {\bibfnamefont {G.}~\bibnamefont {Meneghini}}, \bibinfo {author}
  {\bibfnamefont {A.}~\bibnamefont {AlMutairi}}, \bibinfo {author}
  {\bibfnamefont {K.}~\bibnamefont {Watanabe}}, \bibinfo {author}
  {\bibfnamefont {T.}~\bibnamefont {Taniguchi}}, \bibinfo {author}
  {\bibfnamefont {D.}~\bibnamefont {Steil}}, \bibinfo {author} {\bibfnamefont
  {S.}~\bibnamefont {Steil}}, \bibinfo {author} {\bibfnamefont {R.~T.}\
  \bibnamefont {Weitz}}, \bibinfo {author} {\bibfnamefont {G.~S.~M.}\
  \bibnamefont {Jansen}}, \bibinfo {author} {\bibfnamefont {S.}~\bibnamefont
  {Hofmann}}, \bibinfo {author} {\bibfnamefont {S.}~\bibnamefont {Brem}},
  \bibinfo {author} {\bibfnamefont {E.}~\bibnamefont {Malic}}, \bibinfo
  {author} {\bibfnamefont {M.}~\bibnamefont {Reutzel}},\ and\ \bibinfo {author}
  {\bibfnamefont {S.}~\bibnamefont {Mathias}},\ }\bibfield  {title} {\bibinfo
  {title} {Probing electron-hole coulomb correlations in the exciton landscape
  of a twisted semiconductor heterostructure},\ }\href
  {https://doi.org/doi:10.1126/sciadv.adi1323} {\bibfield  {journal} {\bibinfo
  {journal} {Science Advances}\ }\textbf {\bibinfo {volume} {10}},\ \bibinfo
  {pages} {eadi1323} (\bibinfo {year} {2024})}\BibitemShut {NoStop}%
\bibitem [{\citenamefont {Schmitt}\ \emph {et~al.}(2025)\citenamefont
  {Schmitt}, \citenamefont {Bange}, \citenamefont {Bennecke}, \citenamefont
  {Meneghini}, \citenamefont {AlMutairi}, \citenamefont {Merboldt},
  \citenamefont {Pöhls}, \citenamefont {Watanabe}, \citenamefont {Taniguchi},
  \citenamefont {Steil}, \citenamefont {Steil}, \citenamefont {Weitz},
  \citenamefont {Hofmann}, \citenamefont {Brem}, \citenamefont {Jansen},
  \citenamefont {Malic}, \citenamefont {Mathias},\ and\ \citenamefont
  {Reutzel}}]{Schmitt25natpho}%
  \BibitemOpen
  \bibfield  {author} {\bibinfo {author} {\bibfnamefont {D.}~\bibnamefont
  {Schmitt}}, \bibinfo {author} {\bibfnamefont {J.~P.}\ \bibnamefont {Bange}},
  \bibinfo {author} {\bibfnamefont {W.}~\bibnamefont {Bennecke}}, \bibinfo
  {author} {\bibfnamefont {G.}~\bibnamefont {Meneghini}}, \bibinfo {author}
  {\bibfnamefont {A.}~\bibnamefont {AlMutairi}}, \bibinfo {author}
  {\bibfnamefont {M.}~\bibnamefont {Merboldt}}, \bibinfo {author}
  {\bibfnamefont {J.}~\bibnamefont {Pöhls}}, \bibinfo {author} {\bibfnamefont
  {K.}~\bibnamefont {Watanabe}}, \bibinfo {author} {\bibfnamefont
  {T.}~\bibnamefont {Taniguchi}}, \bibinfo {author} {\bibfnamefont
  {S.}~\bibnamefont {Steil}}, \bibinfo {author} {\bibfnamefont
  {D.}~\bibnamefont {Steil}}, \bibinfo {author} {\bibfnamefont {R.~T.}\
  \bibnamefont {Weitz}}, \bibinfo {author} {\bibfnamefont {S.}~\bibnamefont
  {Hofmann}}, \bibinfo {author} {\bibfnamefont {S.}~\bibnamefont {Brem}},
  \bibinfo {author} {\bibfnamefont {G.~S.~M.}\ \bibnamefont {Jansen}}, \bibinfo
  {author} {\bibfnamefont {E.}~\bibnamefont {Malic}}, \bibinfo {author}
  {\bibfnamefont {S.}~\bibnamefont {Mathias}},\ and\ \bibinfo {author}
  {\bibfnamefont {M.}~\bibnamefont {Reutzel}},\ }\bibfield  {title} {\bibinfo
  {title} {Ultrafast nano-imaging of dark excitons},\ }\bibfield  {journal}
  {\bibinfo  {journal} {Nature Photonics}\ }\href
  {https://doi.org/10.1038/s41566-024-01568-y} {10.1038/s41566-024-01568-y}
  (\bibinfo {year} {2025})\BibitemShut {NoStop}%
\bibitem [{\citenamefont {{Bennecke}}\ \emph {et~al.}(2024)\citenamefont
  {{Bennecke}}, \citenamefont {{Gonzalez Oliva}}, \citenamefont {{Bange}},
  \citenamefont {{Werner}}, \citenamefont {{Schmitt}}, \citenamefont
  {{Merboldt}}, \citenamefont {{Seiler}}, \citenamefont {{Watanabe}},
  \citenamefont {{Taniguchi}}, \citenamefont {{Steil}}, \citenamefont
  {{Weitz}}, \citenamefont {{Puschnig}}, \citenamefont {{Draxl}}, \citenamefont
  {{Matthijs Jansen}}, \citenamefont {{Reutzel}},\ and\ \citenamefont
  {{Mathias}}}]{Bennecke24arxiv}%
  \BibitemOpen
  \bibfield  {author} {\bibinfo {author} {\bibfnamefont {W.}~\bibnamefont
  {{Bennecke}}}, \bibinfo {author} {\bibfnamefont {I.}~\bibnamefont {{Gonzalez
  Oliva}}}, \bibinfo {author} {\bibfnamefont {J.~P.}\ \bibnamefont {{Bange}}},
  \bibinfo {author} {\bibfnamefont {P.}~\bibnamefont {{Werner}}}, \bibinfo
  {author} {\bibfnamefont {D.}~\bibnamefont {{Schmitt}}}, \bibinfo {author}
  {\bibfnamefont {M.}~\bibnamefont {{Merboldt}}}, \bibinfo {author}
  {\bibfnamefont {A.~M.}\ \bibnamefont {{Seiler}}}, \bibinfo {author}
  {\bibfnamefont {K.}~\bibnamefont {{Watanabe}}}, \bibinfo {author}
  {\bibfnamefont {T.}~\bibnamefont {{Taniguchi}}}, \bibinfo {author}
  {\bibfnamefont {D.}~\bibnamefont {{Steil}}}, \bibinfo {author} {\bibfnamefont
  {R.~T.}\ \bibnamefont {{Weitz}}}, \bibinfo {author} {\bibfnamefont
  {P.}~\bibnamefont {{Puschnig}}}, \bibinfo {author} {\bibfnamefont
  {C.}~\bibnamefont {{Draxl}}}, \bibinfo {author} {\bibfnamefont {G.~S.}\
  \bibnamefont {{Matthijs Jansen}}}, \bibinfo {author} {\bibfnamefont
  {M.}~\bibnamefont {{Reutzel}}},\ and\ \bibinfo {author} {\bibfnamefont
  {S.}~\bibnamefont {{Mathias}}},\ }\bibfield  {title} {\bibinfo {title}
  {{Hybrid Frenkel-Wannier excitons facilitate ultrafast energy transfer at a
  2D-organic interface}},\ }\href {https://doi.org/10.48550/arXiv.2411.14993}
  {\bibfield  {journal} {\bibinfo  {journal} {arXiv e-prints}\ ,\ \bibinfo
  {eid} {arXiv:2411.14993}} (\bibinfo {year} {2024})},\ \Eprint
  {https://arxiv.org/abs/2411.14993} {arXiv:2411.14993 [cond-mat.mes-hall]}
  \BibitemShut {NoStop}%
\bibitem [{\citenamefont {Perfetto}\ \emph {et~al.}(2016)\citenamefont
  {Perfetto}, \citenamefont {Sangalli}, \citenamefont {Marini},\ and\
  \citenamefont {Stefanucci}}]{Perfetto16prb}%
  \BibitemOpen
  \bibfield  {author} {\bibinfo {author} {\bibfnamefont {E.}~\bibnamefont
  {Perfetto}}, \bibinfo {author} {\bibfnamefont {D.}~\bibnamefont {Sangalli}},
  \bibinfo {author} {\bibfnamefont {A.}~\bibnamefont {Marini}},\ and\ \bibinfo
  {author} {\bibfnamefont {G.}~\bibnamefont {Stefanucci}},\ }\bibfield  {title}
  {\bibinfo {title} {First-principles approach to excitons in time-resolved and
  angle-resolved photoemission spectra},\ }\href
  {https://doi.org/10.1103/PhysRevB.94.245303} {\bibfield  {journal} {\bibinfo
  {journal} {Physical Review B}\ }\textbf {\bibinfo {volume} {94}},\ \bibinfo
  {pages} {245303} (\bibinfo {year} {2016})}\BibitemShut {NoStop}%
\bibitem [{\citenamefont {Steinhoff}\ \emph {et~al.}(2017)\citenamefont
  {Steinhoff}, \citenamefont {Florian}, \citenamefont {Rösner}, \citenamefont
  {Schönhoff}, \citenamefont {Wehling},\ and\ \citenamefont
  {Jahnke}}]{Steinhoff2017natcom}%
  \BibitemOpen
  \bibfield  {author} {\bibinfo {author} {\bibfnamefont {A.}~\bibnamefont
  {Steinhoff}}, \bibinfo {author} {\bibfnamefont {M.}~\bibnamefont {Florian}},
  \bibinfo {author} {\bibfnamefont {M.}~\bibnamefont {Rösner}}, \bibinfo
  {author} {\bibfnamefont {G.}~\bibnamefont {Schönhoff}}, \bibinfo {author}
  {\bibfnamefont {T.~O.}\ \bibnamefont {Wehling}},\ and\ \bibinfo {author}
  {\bibfnamefont {F.}~\bibnamefont {Jahnke}},\ }\bibfield  {title} {\bibinfo
  {title} {{Exciton fission in monolayer transition metal dichalcogenide
  semiconductors}},\ }\href {https://doi.org/10.1038/s41467-017-01298-6}
  {\bibfield  {journal} {\bibinfo  {journal} {Nature Communications}\ }\textbf
  {\bibinfo {volume} {8}},\ \bibinfo {pages} {1166} (\bibinfo {year}
  {2017})}\BibitemShut {NoStop}%
\bibitem [{\citenamefont {Rustagi}\ and\ \citenamefont
  {Kemper}(2018)}]{Rustagi18prb}%
  \BibitemOpen
  \bibfield  {author} {\bibinfo {author} {\bibfnamefont {A.}~\bibnamefont
  {Rustagi}}\ and\ \bibinfo {author} {\bibfnamefont {A.~F.}\ \bibnamefont
  {Kemper}},\ }\bibfield  {title} {\bibinfo {title} {Photoemission signature of
  excitons},\ }\href {https://doi.org/10.1103/PhysRevB.97.235310} {\bibfield
  {journal} {\bibinfo  {journal} {Physical Review B}\ }\textbf {\bibinfo
  {volume} {97}},\ \bibinfo {pages} {235310} (\bibinfo {year}
  {2018})}\BibitemShut {NoStop}%
\bibitem [{\citenamefont {Christiansen}\ \emph {et~al.}(2019)\citenamefont
  {Christiansen}, \citenamefont {Selig}, \citenamefont {Malic}, \citenamefont
  {Ernstorfer},\ and\ \citenamefont {Knorr}}]{Christiansen19prb}%
  \BibitemOpen
  \bibfield  {author} {\bibinfo {author} {\bibfnamefont {D.}~\bibnamefont
  {Christiansen}}, \bibinfo {author} {\bibfnamefont {M.}~\bibnamefont {Selig}},
  \bibinfo {author} {\bibfnamefont {E.}~\bibnamefont {Malic}}, \bibinfo
  {author} {\bibfnamefont {R.}~\bibnamefont {Ernstorfer}},\ and\ \bibinfo
  {author} {\bibfnamefont {A.}~\bibnamefont {Knorr}},\ }\bibfield  {title}
  {\bibinfo {title} {Theory of exciton dynamics in time-resolved {ARPES}:
  {Intra}- and intervalley scattering in two-dimensional semiconductors},\
  }\href {https://doi.org/10.1103/PhysRevB.100.205401} {\bibfield  {journal}
  {\bibinfo  {journal} {Physical Review B}\ }\textbf {\bibinfo {volume}
  {100}},\ \bibinfo {pages} {205401} (\bibinfo {year} {2019})}\BibitemShut
  {NoStop}%
\bibitem [{\citenamefont {Mak}\ \emph {et~al.}(2010)\citenamefont {Mak},
  \citenamefont {Lee}, \citenamefont {Hone}, \citenamefont {Shan},\ and\
  \citenamefont {Heinz}}]{Mak10prl}%
  \BibitemOpen
  \bibfield  {author} {\bibinfo {author} {\bibfnamefont {K.~F.}\ \bibnamefont
  {Mak}}, \bibinfo {author} {\bibfnamefont {C.}~\bibnamefont {Lee}}, \bibinfo
  {author} {\bibfnamefont {J.}~\bibnamefont {Hone}}, \bibinfo {author}
  {\bibfnamefont {J.}~\bibnamefont {Shan}},\ and\ \bibinfo {author}
  {\bibfnamefont {T.~F.}\ \bibnamefont {Heinz}},\ }\bibfield  {title} {\bibinfo
  {title} {Atomically thin {${\mathrm{MoS}}_{2}$}: A new direct-gap
  semiconductor},\ }\href {https://doi.org/10.1103/PhysRevLett.105.136805}
  {\bibfield  {journal} {\bibinfo  {journal} {Physical Review Letters}\
  }\textbf {\bibinfo {volume} {105}},\ \bibinfo {pages} {136805} (\bibinfo
  {year} {2010})}\BibitemShut {NoStop}%
\bibitem [{\citenamefont {Scheuschner}\ \emph {et~al.}(2014)\citenamefont
  {Scheuschner}, \citenamefont {Ochedowski}, \citenamefont {Kaulitz},
  \citenamefont {Gillen}, \citenamefont {Schleberger},\ and\ \citenamefont
  {Maultzsch}}]{Scheuschner14prb}%
  \BibitemOpen
  \bibfield  {author} {\bibinfo {author} {\bibfnamefont {N.}~\bibnamefont
  {Scheuschner}}, \bibinfo {author} {\bibfnamefont {O.}~\bibnamefont
  {Ochedowski}}, \bibinfo {author} {\bibfnamefont {A.-M.}\ \bibnamefont
  {Kaulitz}}, \bibinfo {author} {\bibfnamefont {R.}~\bibnamefont {Gillen}},
  \bibinfo {author} {\bibfnamefont {M.}~\bibnamefont {Schleberger}},\ and\
  \bibinfo {author} {\bibfnamefont {J.}~\bibnamefont {Maultzsch}},\ }\bibfield
  {title} {\bibinfo {title} {Photoluminescence of freestanding single- and
  few-layer {MoS}$_2$},\ }\href {https://doi.org/10.1103/PhysRevB.89.125406}
  {\bibfield  {journal} {\bibinfo  {journal} {Physical Review B}\ }\textbf
  {\bibinfo {volume} {89}},\ \bibinfo {pages} {125406} (\bibinfo {year}
  {2014})}\BibitemShut {NoStop}%
\bibitem [{\citenamefont {Keunecke}\ \emph {et~al.}(2020)\citenamefont
  {Keunecke}, \citenamefont {Möller}, \citenamefont {Schmitt}, \citenamefont
  {Nolte}, \citenamefont {Jansen}, \citenamefont {Reutzel}, \citenamefont
  {Gutberlet}, \citenamefont {Halasi}, \citenamefont {Steil}, \citenamefont
  {Steil},\ and\ \citenamefont {Mathias}}]{Keunecke20timeresolved}%
  \BibitemOpen
  \bibfield  {author} {\bibinfo {author} {\bibfnamefont {M.}~\bibnamefont
  {Keunecke}}, \bibinfo {author} {\bibfnamefont {C.}~\bibnamefont {Möller}},
  \bibinfo {author} {\bibfnamefont {D.}~\bibnamefont {Schmitt}}, \bibinfo
  {author} {\bibfnamefont {H.}~\bibnamefont {Nolte}}, \bibinfo {author}
  {\bibfnamefont {G.~S.~M.}\ \bibnamefont {Jansen}}, \bibinfo {author}
  {\bibfnamefont {M.}~\bibnamefont {Reutzel}}, \bibinfo {author} {\bibfnamefont
  {M.}~\bibnamefont {Gutberlet}}, \bibinfo {author} {\bibfnamefont
  {G.}~\bibnamefont {Halasi}}, \bibinfo {author} {\bibfnamefont
  {D.}~\bibnamefont {Steil}}, \bibinfo {author} {\bibfnamefont
  {S.}~\bibnamefont {Steil}},\ and\ \bibinfo {author} {\bibfnamefont
  {S.}~\bibnamefont {Mathias}},\ }\bibfield  {title} {\bibinfo {title}
  {{Time-resolved momentum microscopy with a 1 {MHz} high-harmonic extreme
  ultraviolet beamline}},\ }\href {https://doi.org/10.1063/5.0006531}
  {\bibfield  {journal} {\bibinfo  {journal} {Review of Scientific
  Instruments}\ }\textbf {\bibinfo {volume} {91}},\ \bibinfo {pages} {063905}
  (\bibinfo {year} {2020})}\BibitemShut {NoStop}%
\bibitem [{\citenamefont {Medjanik}\ \emph {et~al.}(2017)\citenamefont
  {Medjanik}, \citenamefont {Fedchenko}, \citenamefont {Chernov}, \citenamefont
  {Kutnyakhov}, \citenamefont {Ellguth}, \citenamefont {Oelsner}, \citenamefont
  {Sch{\"o}nhense}, \citenamefont {Peixoto}, \citenamefont {Lutz},
  \citenamefont {Min}, \citenamefont {Reinert}, \citenamefont {D{\"a}ster},
  \citenamefont {Acremann}, \citenamefont {Viefhaus}, \citenamefont {Wurth},
  \citenamefont {Elmers},\ and\ \citenamefont
  {Sch{\"o}nhense}}]{medjanik_direct_2017}%
  \BibitemOpen
  \bibfield  {author} {\bibinfo {author} {\bibfnamefont {K.}~\bibnamefont
  {Medjanik}}, \bibinfo {author} {\bibfnamefont {O.}~\bibnamefont {Fedchenko}},
  \bibinfo {author} {\bibfnamefont {S.}~\bibnamefont {Chernov}}, \bibinfo
  {author} {\bibfnamefont {D.}~\bibnamefont {Kutnyakhov}}, \bibinfo {author}
  {\bibfnamefont {M.}~\bibnamefont {Ellguth}}, \bibinfo {author} {\bibfnamefont
  {A.}~\bibnamefont {Oelsner}}, \bibinfo {author} {\bibfnamefont
  {B.}~\bibnamefont {Sch{\"o}nhense}}, \bibinfo {author} {\bibfnamefont
  {T.~R.~F.}\ \bibnamefont {Peixoto}}, \bibinfo {author} {\bibfnamefont
  {P.}~\bibnamefont {Lutz}}, \bibinfo {author} {\bibfnamefont {C.-H.}\
  \bibnamefont {Min}}, \bibinfo {author} {\bibfnamefont {F.}~\bibnamefont
  {Reinert}}, \bibinfo {author} {\bibfnamefont {S.}~\bibnamefont {D{\"a}ster}},
  \bibinfo {author} {\bibfnamefont {Y.}~\bibnamefont {Acremann}}, \bibinfo
  {author} {\bibfnamefont {J.}~\bibnamefont {Viefhaus}}, \bibinfo {author}
  {\bibfnamefont {W.}~\bibnamefont {Wurth}}, \bibinfo {author} {\bibfnamefont
  {H.~J.}\ \bibnamefont {Elmers}},\ and\ \bibinfo {author} {\bibfnamefont
  {G.}~\bibnamefont {Sch{\"o}nhense}},\ }\bibfield  {title} {\bibinfo {title}
  {{Direct {3D} mapping of the Fermi surface and Fermi velocity}},\ }\href
  {https://doi.org/10.1038/nmat4875} {\bibfield  {journal} {\bibinfo  {journal}
  {Nature Materials}\ }\textbf {\bibinfo {volume} {16}},\ \bibinfo {pages}
  {615} (\bibinfo {year} {2017})}\BibitemShut {NoStop}%
\bibitem [{\citenamefont {Meneghini}\ \emph
  {et~al.}(2022{\natexlab{a}})\citenamefont {Meneghini}, \citenamefont {Brem},\
  and\ \citenamefont {Malic}}]{Meneghini22naturalsciences}%
  \BibitemOpen
  \bibfield  {author} {\bibinfo {author} {\bibfnamefont {G.}~\bibnamefont
  {Meneghini}}, \bibinfo {author} {\bibfnamefont {S.}~\bibnamefont {Brem}},\
  and\ \bibinfo {author} {\bibfnamefont {E.}~\bibnamefont {Malic}},\ }\bibfield
   {title} {\bibinfo {title} {Ultrafast phonon-driven charge transfer in {van
  der Waals} heterostructures},\ }\href
  {https://doi.org/https://doi.org/10.1002/ntls.20220014} {\bibfield  {journal}
  {\bibinfo  {journal} {Natural Sciences}\ }\textbf {\bibinfo {volume} {2}},\
  \bibinfo {pages} {e20220014} (\bibinfo {year}
  {2022}{\natexlab{a}})}\BibitemShut {NoStop}%
\bibitem [{\citenamefont {Jin}\ \emph {et~al.}(2014{\natexlab{a}})\citenamefont
  {Jin}, \citenamefont {Li}, \citenamefont {Mullen},\ and\ \citenamefont
  {Kim}}]{Jin14prb}%
  \BibitemOpen
  \bibfield  {author} {\bibinfo {author} {\bibfnamefont {Z.}~\bibnamefont
  {Jin}}, \bibinfo {author} {\bibfnamefont {X.}~\bibnamefont {Li}}, \bibinfo
  {author} {\bibfnamefont {J.~T.}\ \bibnamefont {Mullen}},\ and\ \bibinfo
  {author} {\bibfnamefont {K.~W.}\ \bibnamefont {Kim}},\ }\bibfield  {title}
  {\bibinfo {title} {Intrinsic transport properties of electrons and holes in
  monolayer transition-metal dichalcogenides},\ }\href
  {https://doi.org/10.1103/PhysRevB.90.045422} {\bibfield  {journal} {\bibinfo
  {journal} {Physical Review B}\ }\textbf {\bibinfo {volume} {90}},\ \bibinfo
  {pages} {045422} (\bibinfo {year} {2014}{\natexlab{a}})}\BibitemShut
  {NoStop}%
\bibitem [{\citenamefont {Jin}\ \emph {et~al.}(2013)\citenamefont {Jin},
  \citenamefont {Yeh}, \citenamefont {Zaki}, \citenamefont {Zhang},
  \citenamefont {Sadowski}, \citenamefont {Al-Mahboob}, \citenamefont {van~der
  Zande}, \citenamefont {Chenet}, \citenamefont {Dadap}, \citenamefont
  {Herman}, \citenamefont {Sutter}, \citenamefont {Hone},\ and\ \citenamefont
  {Osgood}}]{Wencan13prl}%
  \BibitemOpen
  \bibfield  {author} {\bibinfo {author} {\bibfnamefont {W.}~\bibnamefont
  {Jin}}, \bibinfo {author} {\bibfnamefont {P.-C.}\ \bibnamefont {Yeh}},
  \bibinfo {author} {\bibfnamefont {N.}~\bibnamefont {Zaki}}, \bibinfo {author}
  {\bibfnamefont {D.}~\bibnamefont {Zhang}}, \bibinfo {author} {\bibfnamefont
  {J.~T.}\ \bibnamefont {Sadowski}}, \bibinfo {author} {\bibfnamefont
  {A.}~\bibnamefont {Al-Mahboob}}, \bibinfo {author} {\bibfnamefont {A.~M.}\
  \bibnamefont {van~der Zande}}, \bibinfo {author} {\bibfnamefont {D.~A.}\
  \bibnamefont {Chenet}}, \bibinfo {author} {\bibfnamefont {J.~I.}\
  \bibnamefont {Dadap}}, \bibinfo {author} {\bibfnamefont {I.~P.}\ \bibnamefont
  {Herman}}, \bibinfo {author} {\bibfnamefont {P.}~\bibnamefont {Sutter}},
  \bibinfo {author} {\bibfnamefont {J.}~\bibnamefont {Hone}},\ and\ \bibinfo
  {author} {\bibfnamefont {R.~M.}\ \bibnamefont {Osgood}},\ }\bibfield  {title}
  {\bibinfo {title} {Direct measurement of the thickness-dependent electronic
  band structure of {MoS$_2$} using angle-resolved photoemission
  spectroscopy},\ }\href
  {http://link.aps.org/doi/10.1103/PhysRevLett.111.106801} {\bibfield
  {journal} {\bibinfo  {journal} {Physical Review Letters}\ }\textbf {\bibinfo
  {volume} {111}},\ \bibinfo {pages} {106801} (\bibinfo {year}
  {2013})}\BibitemShut {NoStop}%
\bibitem [{\citenamefont {Bennecke}\ \emph {et~al.}(2024)\citenamefont
  {Bennecke}, \citenamefont {Windischbacher}, \citenamefont {Schmitt},
  \citenamefont {Bange}, \citenamefont {Hemm}, \citenamefont {Kern},
  \citenamefont {D’Avino}, \citenamefont {Blase}, \citenamefont {Steil},
  \citenamefont {Steil}, \citenamefont {Aeschlimann}, \citenamefont
  {Stadtmüller}, \citenamefont {Reutzel}, \citenamefont {Puschnig},
  \citenamefont {Jansen},\ and\ \citenamefont {Mathias}}]{Bennecke24natcom}%
  \BibitemOpen
  \bibfield  {author} {\bibinfo {author} {\bibfnamefont {W.}~\bibnamefont
  {Bennecke}}, \bibinfo {author} {\bibfnamefont {A.}~\bibnamefont
  {Windischbacher}}, \bibinfo {author} {\bibfnamefont {D.}~\bibnamefont
  {Schmitt}}, \bibinfo {author} {\bibfnamefont {J.~P.}\ \bibnamefont {Bange}},
  \bibinfo {author} {\bibfnamefont {R.}~\bibnamefont {Hemm}}, \bibinfo {author}
  {\bibfnamefont {C.~S.}\ \bibnamefont {Kern}}, \bibinfo {author}
  {\bibfnamefont {G.}~\bibnamefont {D’Avino}}, \bibinfo {author}
  {\bibfnamefont {X.}~\bibnamefont {Blase}}, \bibinfo {author} {\bibfnamefont
  {D.}~\bibnamefont {Steil}}, \bibinfo {author} {\bibfnamefont
  {S.}~\bibnamefont {Steil}}, \bibinfo {author} {\bibfnamefont
  {M.}~\bibnamefont {Aeschlimann}}, \bibinfo {author} {\bibfnamefont
  {B.}~\bibnamefont {Stadtmüller}}, \bibinfo {author} {\bibfnamefont
  {M.}~\bibnamefont {Reutzel}}, \bibinfo {author} {\bibfnamefont
  {P.}~\bibnamefont {Puschnig}}, \bibinfo {author} {\bibfnamefont {G.~S.~M.}\
  \bibnamefont {Jansen}},\ and\ \bibinfo {author} {\bibfnamefont
  {S.}~\bibnamefont {Mathias}},\ }\bibfield  {title} {\bibinfo {title}
  {Disentangling the multiorbital contributions of excitons by photoemission
  exciton tomography},\ }\href {https://doi.org/10.1038/s41467-024-45973-x}
  {\bibfield  {journal} {\bibinfo  {journal} {Nature Communications}\ }\textbf
  {\bibinfo {volume} {15}},\ \bibinfo {pages} {1804} (\bibinfo {year}
  {2024})}\BibitemShut {NoStop}%
\bibitem [{\citenamefont {Fei}\ \emph {et~al.}(2016)\citenamefont {Fei},
  \citenamefont {Scott}, \citenamefont {Gosztola}, \citenamefont {Foley},
  \citenamefont {Yan}, \citenamefont {Mandrus}, \citenamefont {Wen},
  \citenamefont {Zhou}, \citenamefont {Zhang}, \citenamefont {Sun},
  \citenamefont {Guest}, \citenamefont {Gray}, \citenamefont {Bao},
  \citenamefont {Wiederrecht},\ and\ \citenamefont {Xu}}]{Fei16prb}%
  \BibitemOpen
  \bibfield  {author} {\bibinfo {author} {\bibfnamefont {Z.}~\bibnamefont
  {Fei}}, \bibinfo {author} {\bibfnamefont {M.~E.}\ \bibnamefont {Scott}},
  \bibinfo {author} {\bibfnamefont {D.~J.}\ \bibnamefont {Gosztola}}, \bibinfo
  {author} {\bibfnamefont {J.~J.}\ \bibnamefont {Foley}}, \bibinfo {author}
  {\bibfnamefont {J.}~\bibnamefont {Yan}}, \bibinfo {author} {\bibfnamefont
  {D.~G.}\ \bibnamefont {Mandrus}}, \bibinfo {author} {\bibfnamefont
  {H.}~\bibnamefont {Wen}}, \bibinfo {author} {\bibfnamefont {P.}~\bibnamefont
  {Zhou}}, \bibinfo {author} {\bibfnamefont {D.~W.}\ \bibnamefont {Zhang}},
  \bibinfo {author} {\bibfnamefont {Y.}~\bibnamefont {Sun}}, \bibinfo {author}
  {\bibfnamefont {J.~R.}\ \bibnamefont {Guest}}, \bibinfo {author}
  {\bibfnamefont {S.~K.}\ \bibnamefont {Gray}}, \bibinfo {author}
  {\bibfnamefont {W.}~\bibnamefont {Bao}}, \bibinfo {author} {\bibfnamefont
  {G.~P.}\ \bibnamefont {Wiederrecht}},\ and\ \bibinfo {author} {\bibfnamefont
  {X.}~\bibnamefont {Xu}},\ }\bibfield  {title} {\bibinfo {title} {Nano-optical
  imaging of $\mathrm{WS}{\mathrm{e}}_{2}$ waveguide modes revealing
  light-exciton interactions},\ }\href
  {https://doi.org/10.1103/PhysRevB.94.081402} {\bibfield  {journal} {\bibinfo
  {journal} {Physical Review B}\ }\textbf {\bibinfo {volume} {94}},\ \bibinfo
  {pages} {081402} (\bibinfo {year} {2016})}\BibitemShut {NoStop}%
\bibitem [{\citenamefont {Byrnes}\ \emph {et~al.}(2014)\citenamefont {Byrnes},
  \citenamefont {Kim},\ and\ \citenamefont {Yamamoto}}]{Byrnes14natphys}%
  \BibitemOpen
  \bibfield  {author} {\bibinfo {author} {\bibfnamefont {T.}~\bibnamefont
  {Byrnes}}, \bibinfo {author} {\bibfnamefont {N.~Y.}\ \bibnamefont {Kim}},\
  and\ \bibinfo {author} {\bibfnamefont {Y.}~\bibnamefont {Yamamoto}},\
  }\bibfield  {title} {\bibinfo {title} {Exciton–polariton condensates},\
  }\href {https://doi.org/10.1038/nphys3143} {\bibfield  {journal} {\bibinfo
  {journal} {Nature Physics}\ }\textbf {\bibinfo {volume} {10}},\ \bibinfo
  {pages} {803} (\bibinfo {year} {2014})}\BibitemShut {NoStop}%
\bibitem [{\citenamefont {Seiler}\ \emph {et~al.}(2024)\citenamefont {Seiler},
  \citenamefont {Zhumagulov}, \citenamefont {Zollner}, \citenamefont {Yoon},
  \citenamefont {Urbaniak}, \citenamefont {Geisenhof}, \citenamefont
  {Watanabe}, \citenamefont {Taniguchi}, \citenamefont {Fabian}, \citenamefont
  {Zhang},\ and\ \citenamefont {Weitz}}]{seiler2024arxiv}%
  \BibitemOpen
  \bibfield  {author} {\bibinfo {author} {\bibfnamefont {A.~M.}\ \bibnamefont
  {Seiler}}, \bibinfo {author} {\bibfnamefont {Y.}~\bibnamefont {Zhumagulov}},
  \bibinfo {author} {\bibfnamefont {K.}~\bibnamefont {Zollner}}, \bibinfo
  {author} {\bibfnamefont {C.}~\bibnamefont {Yoon}}, \bibinfo {author}
  {\bibfnamefont {D.}~\bibnamefont {Urbaniak}}, \bibinfo {author}
  {\bibfnamefont {F.~R.}\ \bibnamefont {Geisenhof}}, \bibinfo {author}
  {\bibfnamefont {K.}~\bibnamefont {Watanabe}}, \bibinfo {author}
  {\bibfnamefont {T.}~\bibnamefont {Taniguchi}}, \bibinfo {author}
  {\bibfnamefont {J.}~\bibnamefont {Fabian}}, \bibinfo {author} {\bibfnamefont
  {F.}~\bibnamefont {Zhang}},\ and\ \bibinfo {author} {\bibfnamefont {R.~T.}\
  \bibnamefont {Weitz}},\ }\href {https://arxiv.org/abs/2403.17140} {\bibinfo
  {title} {Layer-selective spin-orbit coupling and strong correlation in
  bilayer graphene}} (\bibinfo {year} {2024}),\ \Eprint
  {https://arxiv.org/abs/2403.17140} {arXiv:2403.17140 [cond-mat.mes-hall]}
  \BibitemShut {NoStop}%
\bibitem [{\citenamefont {Düvel}\ \emph {et~al.}(2022)\citenamefont {Düvel},
  \citenamefont {Merboldt}, \citenamefont {Bange}, \citenamefont {Strauch},
  \citenamefont {Stellbrink}, \citenamefont {Pierz}, \citenamefont
  {Schumacher}, \citenamefont {Momeni}, \citenamefont {Steil}, \citenamefont
  {Jansen}, \citenamefont {Steil}, \citenamefont {Novko}, \citenamefont
  {Mathias},\ and\ \citenamefont {Reutzel}}]{Duvel22nanolett}%
  \BibitemOpen
  \bibfield  {author} {\bibinfo {author} {\bibfnamefont {M.}~\bibnamefont
  {Düvel}}, \bibinfo {author} {\bibfnamefont {M.}~\bibnamefont {Merboldt}},
  \bibinfo {author} {\bibfnamefont {J.~P.}\ \bibnamefont {Bange}}, \bibinfo
  {author} {\bibfnamefont {H.}~\bibnamefont {Strauch}}, \bibinfo {author}
  {\bibfnamefont {M.}~\bibnamefont {Stellbrink}}, \bibinfo {author}
  {\bibfnamefont {K.}~\bibnamefont {Pierz}}, \bibinfo {author} {\bibfnamefont
  {H.~W.}\ \bibnamefont {Schumacher}}, \bibinfo {author} {\bibfnamefont
  {D.}~\bibnamefont {Momeni}}, \bibinfo {author} {\bibfnamefont
  {D.}~\bibnamefont {Steil}}, \bibinfo {author} {\bibfnamefont {G.~S.~M.}\
  \bibnamefont {Jansen}}, \bibinfo {author} {\bibfnamefont {S.}~\bibnamefont
  {Steil}}, \bibinfo {author} {\bibfnamefont {D.}~\bibnamefont {Novko}},
  \bibinfo {author} {\bibfnamefont {S.}~\bibnamefont {Mathias}},\ and\ \bibinfo
  {author} {\bibfnamefont {M.}~\bibnamefont {Reutzel}},\ }\bibfield  {title}
  {\bibinfo {title} {Far-from-equilibrium electron–phonon interactions in
  optically excited graphene},\ }\href
  {https://doi.org/10.1021/acs.nanolett.2c01325} {\bibfield  {journal}
  {\bibinfo  {journal} {Nano Letters}\ }\textbf {\bibinfo {volume} {22}},\
  \bibinfo {pages} {4897} (\bibinfo {year} {2022})}\BibitemShut {NoStop}%
\bibitem [{\citenamefont {Merboldt}\ \emph {et~al.}(2025)\citenamefont
  {Merboldt}, \citenamefont {Schüler}, \citenamefont {Schmitt}, \citenamefont
  {Bange}, \citenamefont {Bennecke}, \citenamefont {Gadge}, \citenamefont
  {Pierz}, \citenamefont {Schumacher}, \citenamefont {Momeni}, \citenamefont
  {Steil}, \citenamefont {Manmana}, \citenamefont {Sentef}, \citenamefont
  {Reutzel},\ and\ \citenamefont {Mathias}}]{Merboldt25NatPhys}%
  \BibitemOpen
  \bibfield  {author} {\bibinfo {author} {\bibfnamefont {M.}~\bibnamefont
  {Merboldt}}, \bibinfo {author} {\bibfnamefont {M.}~\bibnamefont {Schüler}},
  \bibinfo {author} {\bibfnamefont {D.}~\bibnamefont {Schmitt}}, \bibinfo
  {author} {\bibfnamefont {J.~P.}\ \bibnamefont {Bange}}, \bibinfo {author}
  {\bibfnamefont {W.}~\bibnamefont {Bennecke}}, \bibinfo {author}
  {\bibfnamefont {K.}~\bibnamefont {Gadge}}, \bibinfo {author} {\bibfnamefont
  {K.}~\bibnamefont {Pierz}}, \bibinfo {author} {\bibfnamefont {H.~W.}\
  \bibnamefont {Schumacher}}, \bibinfo {author} {\bibfnamefont
  {D.}~\bibnamefont {Momeni}}, \bibinfo {author} {\bibfnamefont
  {D.}~\bibnamefont {Steil}}, \bibinfo {author} {\bibfnamefont {S.~R.}\
  \bibnamefont {Manmana}}, \bibinfo {author} {\bibfnamefont {M.~A.}\
  \bibnamefont {Sentef}}, \bibinfo {author} {\bibfnamefont {M.}~\bibnamefont
  {Reutzel}},\ and\ \bibinfo {author} {\bibfnamefont {S.}~\bibnamefont
  {Mathias}},\ }\bibfield  {title} {\bibinfo {title} {Observation of {F}loquet
  states in graphene},\ }\bibfield  {journal} {\bibinfo  {journal} {Nature
  Physics}\ }\href {https://doi.org/10.1038/s41567-025-02889-7}
  {10.1038/s41567-025-02889-7} (\bibinfo {year} {2025})\BibitemShut {NoStop}%
\bibitem [{\citenamefont {Miaja-Avila}\ \emph {et~al.}(2006)\citenamefont
  {Miaja-Avila}, \citenamefont {Lei}, \citenamefont {Aeschlimann},
  \citenamefont {Gland}, \citenamefont {Murnane}, \citenamefont {Kapteyn},\
  and\ \citenamefont {Saathoff}}]{miaja2006laser}%
  \BibitemOpen
  \bibfield  {author} {\bibinfo {author} {\bibfnamefont {L.}~\bibnamefont
  {Miaja-Avila}}, \bibinfo {author} {\bibfnamefont {C.}~\bibnamefont {Lei}},
  \bibinfo {author} {\bibfnamefont {M.}~\bibnamefont {Aeschlimann}}, \bibinfo
  {author} {\bibfnamefont {J.}~\bibnamefont {Gland}}, \bibinfo {author}
  {\bibfnamefont {M.}~\bibnamefont {Murnane}}, \bibinfo {author} {\bibfnamefont
  {H.}~\bibnamefont {Kapteyn}},\ and\ \bibinfo {author} {\bibfnamefont
  {G.}~\bibnamefont {Saathoff}},\ }\bibfield  {title} {\bibinfo {title}
  {Laser-assisted photoelectric effect from surfaces},\ }\href@noop {}
  {\bibfield  {journal} {\bibinfo  {journal} {Physical Review Letters}\
  }\textbf {\bibinfo {volume} {97}},\ \bibinfo {pages} {113604} (\bibinfo
  {year} {2006})}\BibitemShut {NoStop}%
\bibitem [{\citenamefont {Sch{\"o}nhense}\ \emph {et~al.}(2018)\citenamefont
  {Sch{\"o}nhense}, \citenamefont {Medjanik}, \citenamefont {Fedchenko},
  \citenamefont {Chernov}, \citenamefont {Ellguth}, \citenamefont {Vasilyev},
  \citenamefont {Oelsner}, \citenamefont {Viefhaus}, \citenamefont
  {Kutnyakhov}, \citenamefont {Wurth}, \citenamefont {Elmers},\ and\
  \citenamefont {Sch{\"o}nhense}}]{schonhense_multidimensional_2018}%
  \BibitemOpen
  \bibfield  {author} {\bibinfo {author} {\bibfnamefont {B.}~\bibnamefont
  {Sch{\"o}nhense}}, \bibinfo {author} {\bibfnamefont {K.}~\bibnamefont
  {Medjanik}}, \bibinfo {author} {\bibfnamefont {O.}~\bibnamefont {Fedchenko}},
  \bibinfo {author} {\bibfnamefont {S.}~\bibnamefont {Chernov}}, \bibinfo
  {author} {\bibfnamefont {M.}~\bibnamefont {Ellguth}}, \bibinfo {author}
  {\bibfnamefont {D.}~\bibnamefont {Vasilyev}}, \bibinfo {author}
  {\bibfnamefont {A.}~\bibnamefont {Oelsner}}, \bibinfo {author} {\bibfnamefont
  {J.}~\bibnamefont {Viefhaus}}, \bibinfo {author} {\bibfnamefont
  {D.}~\bibnamefont {Kutnyakhov}}, \bibinfo {author} {\bibfnamefont
  {W.}~\bibnamefont {Wurth}}, \bibinfo {author} {\bibfnamefont {H.~J.}\
  \bibnamefont {Elmers}},\ and\ \bibinfo {author} {\bibfnamefont
  {G.}~\bibnamefont {Sch{\"o}nhense}},\ }\bibfield  {title} {\bibinfo {title}
  {Multidimensional photoemission spectroscopy—the space-charge limit},\
  }\href {https://doi.org/10.1088/1367-2630/aaa262} {\bibfield  {journal}
  {\bibinfo  {journal} {New Journal of Physics}\ }\textbf {\bibinfo {volume}
  {20}},\ \bibinfo {pages} {033004} (\bibinfo {year} {2018})}\BibitemShut
  {NoStop}%
\bibitem [{\citenamefont {Roth}\ \emph {et~al.}(2024)\citenamefont {Roth},
  \citenamefont {Mahl}, \citenamefont {Borgwardt}, \citenamefont {Wenthaus},
  \citenamefont {Brausse}, \citenamefont {Garbe}, \citenamefont {Gessner},\
  and\ \citenamefont {Eberhardt}}]{Roth24prl}%
  \BibitemOpen
  \bibfield  {author} {\bibinfo {author} {\bibfnamefont {F.}~\bibnamefont
  {Roth}}, \bibinfo {author} {\bibfnamefont {J.}~\bibnamefont {Mahl}}, \bibinfo
  {author} {\bibfnamefont {M.}~\bibnamefont {Borgwardt}}, \bibinfo {author}
  {\bibfnamefont {L.}~\bibnamefont {Wenthaus}}, \bibinfo {author}
  {\bibfnamefont {F.}~\bibnamefont {Brausse}}, \bibinfo {author} {\bibfnamefont
  {V.}~\bibnamefont {Garbe}}, \bibinfo {author} {\bibfnamefont
  {O.}~\bibnamefont {Gessner}},\ and\ \bibinfo {author} {\bibfnamefont
  {W.}~\bibnamefont {Eberhardt}},\ }\bibfield  {title} {\bibinfo {title}
  {Dynamical nonlinear inversion of the surface photovoltage at {Si}(100)},\
  }\href {https://doi.org/10.1103/PhysRevLett.132.146201} {\bibfield  {journal}
  {\bibinfo  {journal} {Physical Review Letters}\ }\textbf {\bibinfo {volume}
  {132}},\ \bibinfo {pages} {146201} (\bibinfo {year} {2024})}\BibitemShut
  {NoStop}%
\bibitem [{\citenamefont {Li}\ \emph {et~al.}(2014)\citenamefont {Li},
  \citenamefont {Chernikov}, \citenamefont {Zhang}, \citenamefont {Rigosi},
  \citenamefont {Hill}, \citenamefont {van~der Zande}, \citenamefont {Chenet},
  \citenamefont {Shih}, \citenamefont {Hone},\ and\ \citenamefont
  {Heinz}}]{Li14prb}%
  \BibitemOpen
  \bibfield  {author} {\bibinfo {author} {\bibfnamefont {Y.}~\bibnamefont
  {Li}}, \bibinfo {author} {\bibfnamefont {A.}~\bibnamefont {Chernikov}},
  \bibinfo {author} {\bibfnamefont {X.}~\bibnamefont {Zhang}}, \bibinfo
  {author} {\bibfnamefont {A.}~\bibnamefont {Rigosi}}, \bibinfo {author}
  {\bibfnamefont {H.~M.}\ \bibnamefont {Hill}}, \bibinfo {author}
  {\bibfnamefont {A.~M.}\ \bibnamefont {van~der Zande}}, \bibinfo {author}
  {\bibfnamefont {D.~A.}\ \bibnamefont {Chenet}}, \bibinfo {author}
  {\bibfnamefont {E.-M.}\ \bibnamefont {Shih}}, \bibinfo {author}
  {\bibfnamefont {J.}~\bibnamefont {Hone}},\ and\ \bibinfo {author}
  {\bibfnamefont {T.~F.}\ \bibnamefont {Heinz}},\ }\bibfield  {title} {\bibinfo
  {title} {Measurement of the optical dielectric function of monolayer
  transition-metal dichalcogenides:{ ${\mathrm{MoS}}_{2}$,
  $\mathrm{Mo}\mathrm{S}{\mathrm{e}}_{2}$, ${\mathrm{WS}}_{2}$, and
  $\mathrm{WS}{\mathrm{e}}_{2}$}},\ }\href
  {https://doi.org/10.1103/PhysRevB.90.205422} {\bibfield  {journal} {\bibinfo
  {journal} {Physical Review B}\ }\textbf {\bibinfo {volume} {90}},\ \bibinfo
  {pages} {205422} (\bibinfo {year} {2014})}\BibitemShut {NoStop}%
\bibitem [{\citenamefont {Weinelt}\ \emph {et~al.}(2004)\citenamefont
  {Weinelt}, \citenamefont {Kutschera}, \citenamefont {Fauster},\ and\
  \citenamefont {Rohlfing}}]{Weinelt04prl}%
  \BibitemOpen
  \bibfield  {author} {\bibinfo {author} {\bibfnamefont {M.}~\bibnamefont
  {Weinelt}}, \bibinfo {author} {\bibfnamefont {M.}~\bibnamefont {Kutschera}},
  \bibinfo {author} {\bibfnamefont {T.}~\bibnamefont {Fauster}},\ and\ \bibinfo
  {author} {\bibfnamefont {M.}~\bibnamefont {Rohlfing}},\ }\bibfield  {title}
  {\bibinfo {title} {Dynamics of exciton formation at the {Si(100) c(4 x 2)}
  surface},\ }\href {http://link.aps.org/abstract/PRL/v92/e126801} {\bibfield
  {journal} {\bibinfo  {journal} {Phys. Rev. Lett.}\ }\textbf {\bibinfo
  {volume} {92}},\ \bibinfo {pages} {126801} (\bibinfo {year}
  {2004})}\BibitemShut {NoStop}%
\bibitem [{\citenamefont {Brem}\ \emph {et~al.}(2020)\citenamefont {Brem},
  \citenamefont {Lin}, \citenamefont {Gillen}, \citenamefont {Bauer},
  \citenamefont {Maultzsch}, \citenamefont {Lupton},\ and\ \citenamefont
  {Malic}}]{Brem20nanoscale}%
  \BibitemOpen
  \bibfield  {author} {\bibinfo {author} {\bibfnamefont {S.}~\bibnamefont
  {Brem}}, \bibinfo {author} {\bibfnamefont {K.-Q.}\ \bibnamefont {Lin}},
  \bibinfo {author} {\bibfnamefont {R.}~\bibnamefont {Gillen}}, \bibinfo
  {author} {\bibfnamefont {J.~M.}\ \bibnamefont {Bauer}}, \bibinfo {author}
  {\bibfnamefont {J.}~\bibnamefont {Maultzsch}}, \bibinfo {author}
  {\bibfnamefont {J.~M.}\ \bibnamefont {Lupton}},\ and\ \bibinfo {author}
  {\bibfnamefont {E.}~\bibnamefont {Malic}},\ }\bibfield  {title} {\bibinfo
  {title} {Hybridized intervalley moiré excitons and flat bands in twisted
  {WSe$_2$} bilayers},\ }\href {https://doi.org/10.1039/D0NR02160A} {\bibfield
  {journal} {\bibinfo  {journal} {Nanoscale}\ }\textbf {\bibinfo {volume}
  {12}},\ \bibinfo {pages} {11088} (\bibinfo {year} {2020})}\BibitemShut
  {NoStop}%
\bibitem [{\citenamefont {Hagel}\ \emph {et~al.}(2021)\citenamefont {Hagel},
  \citenamefont {Brem}, \citenamefont {Linderälv}, \citenamefont {Erhart},\
  and\ \citenamefont {Malic}}]{Hagel21prr}%
  \BibitemOpen
  \bibfield  {author} {\bibinfo {author} {\bibfnamefont {J.}~\bibnamefont
  {Hagel}}, \bibinfo {author} {\bibfnamefont {S.}~\bibnamefont {Brem}},
  \bibinfo {author} {\bibfnamefont {C.}~\bibnamefont {Linderälv}}, \bibinfo
  {author} {\bibfnamefont {P.}~\bibnamefont {Erhart}},\ and\ \bibinfo {author}
  {\bibfnamefont {E.}~\bibnamefont {Malic}},\ }\bibfield  {title} {\bibinfo
  {title} {Exciton landscape in van der waals heterostructures},\ }\href
  {https://doi.org/10.1103/PhysRevResearch.3.043217} {\bibfield  {journal}
  {\bibinfo  {journal} {Physical Review Research}\ }\textbf {\bibinfo {volume}
  {3}},\ \bibinfo {pages} {043217} (\bibinfo {year} {2021})}\BibitemShut
  {NoStop}%
\bibitem [{\citenamefont {Brem}\ \emph {et~al.}(2018)\citenamefont {Brem},
  \citenamefont {Selig}, \citenamefont {Berghaeuser},\ and\ \citenamefont
  {Malic}}]{Brem18scirep}%
  \BibitemOpen
  \bibfield  {author} {\bibinfo {author} {\bibfnamefont {S.}~\bibnamefont
  {Brem}}, \bibinfo {author} {\bibfnamefont {M.}~\bibnamefont {Selig}},
  \bibinfo {author} {\bibfnamefont {G.}~\bibnamefont {Berghaeuser}},\ and\
  \bibinfo {author} {\bibfnamefont {E.}~\bibnamefont {Malic}},\ }\bibfield
  {title} {\bibinfo {title} {Exciton relaxation cascade in two-dimensional
  transition metal dichalcogenides},\ }\href
  {https://doi.org/10.1038/s41598-018-25906-7} {\bibfield  {journal} {\bibinfo
  {journal} {Scientific Reports}\ }\textbf {\bibinfo {volume} {8}},\ \bibinfo
  {pages} {8238} (\bibinfo {year} {2018})}\BibitemShut {NoStop}%
\bibitem [{\citenamefont {Meneghini}\ \emph
  {et~al.}(2022{\natexlab{b}})\citenamefont {Meneghini}, \citenamefont {Brem},\
  and\ \citenamefont {Malic}}]{meneghini2022ultrafast}%
  \BibitemOpen
  \bibfield  {author} {\bibinfo {author} {\bibfnamefont {G.}~\bibnamefont
  {Meneghini}}, \bibinfo {author} {\bibfnamefont {S.}~\bibnamefont {Brem}},\
  and\ \bibinfo {author} {\bibfnamefont {E.}~\bibnamefont {Malic}},\ }\bibfield
   {title} {\bibinfo {title} {Ultrafast phonon-driven charge transfer in van
  der waals heterostructures},\ }\href@noop {} {\bibfield  {journal} {\bibinfo
  {journal} {Natural Sciences}\ }\textbf {\bibinfo {volume} {2}},\ \bibinfo
  {pages} {e20220014} (\bibinfo {year} {2022}{\natexlab{b}})}\BibitemShut
  {NoStop}%
\bibitem [{\citenamefont {Jin}\ \emph {et~al.}(2014{\natexlab{b}})\citenamefont
  {Jin}, \citenamefont {Li}, \citenamefont {Mullen},\ and\ \citenamefont
  {Kim}}]{PhysRevB.90.045422}%
  \BibitemOpen
  \bibfield  {author} {\bibinfo {author} {\bibfnamefont {Z.}~\bibnamefont
  {Jin}}, \bibinfo {author} {\bibfnamefont {X.}~\bibnamefont {Li}}, \bibinfo
  {author} {\bibfnamefont {J.~T.}\ \bibnamefont {Mullen}},\ and\ \bibinfo
  {author} {\bibfnamefont {K.~W.}\ \bibnamefont {Kim}},\ }\bibfield  {title}
  {\bibinfo {title} {Intrinsic transport properties of electrons and holes in
  monolayer transition-metal dichalcogenides},\ }\href
  {https://doi.org/10.1103/PhysRevB.90.045422} {\bibfield  {journal} {\bibinfo
  {journal} {Physical Review B}\ }\textbf {\bibinfo {volume} {90}},\ \bibinfo
  {pages} {045422} (\bibinfo {year} {2014}{\natexlab{b}})}\BibitemShut
  {NoStop}%
\bibitem [{\citenamefont {Kira}\ and\ \citenamefont
  {Koch}(2006)}]{kira2006many}%
  \BibitemOpen
  \bibfield  {author} {\bibinfo {author} {\bibfnamefont {M.}~\bibnamefont
  {Kira}}\ and\ \bibinfo {author} {\bibfnamefont {S.~W.}\ \bibnamefont
  {Koch}},\ }\bibfield  {title} {\bibinfo {title} {Many-body correlations and
  excitonic effects in semiconductor spectroscopy},\ }\href@noop {} {\bibfield
  {journal} {\bibinfo  {journal} {Progress in quantum electronics}\ }\textbf
  {\bibinfo {volume} {30}},\ \bibinfo {pages} {155} (\bibinfo {year}
  {2006})}\BibitemShut {NoStop}%
\bibitem [{\citenamefont {Haug}\ and\ \citenamefont
  {Koch}(2009)}]{haug2009quantum}%
  \BibitemOpen
  \bibfield  {author} {\bibinfo {author} {\bibfnamefont {H.}~\bibnamefont
  {Haug}}\ and\ \bibinfo {author} {\bibfnamefont {S.~W.}\ \bibnamefont
  {Koch}},\ }\href@noop {} {\emph {\bibinfo {title} {Quantum theory of the
  optical and electronic properties of semiconductors}}}\ (\bibinfo
  {publisher} {World Scientific Publishing Company},\ \bibinfo {year}
  {2009})\BibitemShut {NoStop}%
\bibitem [{\citenamefont {Malic}\ and\ \citenamefont
  {Knorr}(2013)}]{malic2013graphene}%
  \BibitemOpen
  \bibfield  {author} {\bibinfo {author} {\bibfnamefont {E.}~\bibnamefont
  {Malic}}\ and\ \bibinfo {author} {\bibfnamefont {A.}~\bibnamefont {Knorr}},\
  }\href@noop {} {\emph {\bibinfo {title} {Graphene and carbon nanotubes:
  ultrafast optics and relaxation dynamics}}}\ (\bibinfo  {publisher} {John
  Wiley \& Sons},\ \bibinfo {year} {2013})\BibitemShut {NoStop}%
\end{thebibliography}
\end{document}